\documentclass[final,twoside,11pt]{entics}
\usepackage{enticsmacro}
\usepackage{graphicx}
\usepackage{wrapfig}
\usepackage{amsmath,amsbsy,amssymb}
\usepackage{mathtools}
\usepackage[disable]{todonotes}
\newcommand{\hbnote}[1]{\todo[color=green!40]{\scriptsize HB: #1}}
\usepackage[capitalise]{cleveref}
\usepackage{tabularx}
\usepackage{tikz-cd}
\usetikzlibrary{calc}
\usepackage{proof}

\usepackage{enumitem}

\usepackage[appendix=strip,bibliography=common]{apxproof}

\newtheoremrep{theorem}{Theorem}
\newtheoremrep{lemma}[theorem]{Lemma}
\newtheoremrep{proposition}[theorem]{Proposition}
\newtheoremrep{corollary}[theorem]{Corollary}
\theoremstyle{definition}
\newtheoremrep{definition}[theorem]{Definition}
\newtheoremrep{example}[theorem]{Example}
\newtheoremrep{remark}[theorem]{Remark}
\Crefname{lemma}{Lemma}{Lemmas}
\Crefname{proposition}{Proposition}{Propositions}
\Crefname{corollary}{Corollary}{Corollaries}
\Crefname{definition}{Definition}{Definitions}
\Crefname{example}{Example}{Examples}
\Crefname{remark}{Remark}{Remarks}

\newcommand{\power}{\mathcal P}
\newcommand{\cat}[1]{\ensuremath{\mathbf{#1}}}
\newcommand{\set}{\cat{Set}}

\newcommand{\op}[1]{{#1}^{\text{op}}}
\newcommand{\id}[1]{\operatorname{id}_{#1}}
\newcommand{\klcat}{\mathbf {K}{\boldsymbol\ell}}
\newcommand{\gmonad}[1]{{#1}^G}
\newcommand{\inv}[1]{#1^{-1}}
\newcommand{\condset}{\mathbb K}
\newcommand{\cppo}{\cat{Cppo}}

\newcommand{\pos}{\cat{Pos}}
\newcommand{\proj}{\operatorname{pr}}

\newcommand{\inObj}[1]{{\boldsymbol{\in}_{#1}}}

\newcommand{\nat}{\mathbb N}
\newcommand{\fseq}[1]{{#1}^\star}

\newcommand{\kllaw}{$\mathbf{K}{\boldsymbol\ell}$-law}
\newcommand{\kllaws}{$\mathbf{K}{\boldsymbol\ell}$-laws}
\newcommand{\down}{\power_\downarrow}
\newcommand{\history}[1]{\mathord{\downarrow}{#1}}

\newcommand\diagOp{\mathbin\vartriangle}
\newcommand\coDiagOp{\mathbin\triangledown}

\makeatletter
\newcommand{\eqnum}{\refstepcounter{equation}\textup{\tagform@{\theequation}}}
\makeatother

\newcommand{\steps}[1]{%
  \xrightarrow{#1}\mathrel{\mkern-14mu}\rightarrow
}

\newcommand\refrel[3][]{\stackrel{\mathllap{\text{#1\cref{#2}}}} {#3} }

\def\conf{MFPS 2024}
\def\myconf{mfps40} 	%%Fill in the acronym for your conference (with year)
\volume{4}			%Fill in the ENTICS volume number here
\def\volu{4}			% and here
\def\papno{16}			%Fill in your paper number here
\def\mydoi{14751}%
\def\lastname{Luckhardt, Beohar \& K\"upper} %Lastnames appear in the running header
\def\runtitle{On Kleisli liftings} %Short title appears in the running header on even pages.
\def\copynames{H. Beohar, S. K\"upper, D. Luckhardt}    %% Fill in the first initial and last name of the authors
\def\copyname{}
\begin{document}
%%%Note the beginning and end of the frontmatter section that starts here%%%%%
\begin{frontmatter}
 \title{On Kleisli Liftings and Decorated Trace Semantics}
 % \title{On Kleisli liftings and decorated trace semantics\thanksref{ALL}} 						%%Title here and the
 %\thanks[ALL]{EPSRC.}   %%Text of \thanks[ALL} here..
 %\thanks[ALL]{This is an open access article under the CC BY-NC-ND 4.0 license (\href{https://creativecommons.org/licenses/by-nc-nd/4.0/}{creativecommons.org/licenses/by-nc-nd/4.0/}).}
 %%%%%%%%%%%%%%%%%%%%%%%%%%%%			This Thanks is optional.
  %%%%Now the author(s) names(s)%%%%%
\author{Daniel Luckhardt\thanksref{a}\thanksref{emailDaniel}}
\author{Harsh Beohar\thanksref{a}\thanksref{emailHarsh}}%%Note NO SPACE between
\author{Sebastian K\"upper\thanksref{b}\thanksref{emailSebastian}}
%last name and
   		%last name and \thanksref{...}
    %%%Next come the addresses%%%%
   \address[a]{Department of Computer Science\\ University of Sheffield\\				%or between \thanksrefs...
    Sheffield, United Kingdom}
     \thanks[emailDaniel]{Email:  \href{mailto:d.luckhardt@sheffield.ac.uk} {\texttt{\normalshape
    d.luckhardt@sheffield.ac.uk}}.}
    %Supported by EPSRC NIA Grant EP/X019373/1. 							
   \thanks[emailHarsh]{Email: \href{mailto:h.beohar@sheffield.ac.uk} {\texttt{\normalshape
        h.beohar@sheffield.ac.uk}}. }
   %%%Note: if both authors share same institution, only list the address once, after the second
   %%%author.
   %%%There also is a link from the first author to the co-author's address to show how to list
   %%%affiliations to more than one institution, when needed.
  \address[b]{Fakult\"at f\"ur Mathematik und Informatik\\FernUniversit\"at Hagen, \\
    Hagen, Germany}
  \thanks[emailSebastian]{Email:  \href{mailto:sebastian.kuepper@fernuni-hagen.de} {\texttt{\normalshape
        sebastian.kuepper@fernuni-hagen.de}}}

\begin{abstract}
It is well known that Kleisli categories provide a natural language to model side effects. For instance, in the theory of coalgebras, behavioural equivalence coincides with language equivalence (instead of bisimilarity) when  nondeterministic automata are modelled as coalgebras living in the Kleisli category of the powerset monad. In this paper, our aim is to establish decorated trace semantics based on language and ready equivalences for conditional transition systems (CTSs) with/without upgrades. To this end, we model CTSs as coalgebras living in the Kleisli category of a relative monad. Our results are twofold. First, we reduce the problem of defining a Kleisli lifting for the machine endofunctor in the context of a relative monad to the classical notion of Kleisli lifting. Second, we provide a recipe based on indexed categories to construct a Kleisli lifting for general endofunctors.
\end{abstract}
\end{frontmatter}

\section{Introduction}\label{sec:intro}

Coalgebras \cite{RUTTEN2003} are a categorical generalisation of labelled transition systems (LTSs) and state-based systems in general, where the branching type is parameterised by an endofunctor over a category. Coalgebra homomorphisms between any two coalgebras are behaviour preserving maps between the underlying sets of states; often they correspond to some form of functional bisimulations.  Under certain restrictions---for instance, when the underlying endofunctor  over the category $\set$ of sets is bounded \cite{RUTTEN2003}---the final coalgebra exists which can be seen as a universe of all coalgebras of the same type.

As coalgebra homomorphims in the category $\set$ of sets correspond to functional bisimulations, the behavioural equivalence induced by the unique coalegbra homomorphism into the final coalgebra coincides with some form of bisimilarity. Nevertheless, there are many interesting notions of behavioural equivalences other than bisimilarity; for instance, decorated trace equivalences (like trace/language/failure/ready equivalences) on the states of an LTS (see the linear time-branching time spectrum \cite{g:linear-branching-time}).

The clue to get coarser notions than bisimilarity is to consider $TB$-coalgebras with side effects, where $T$ is a monad modelling the implicit side effects (like the powerset monad) and $B=\mathbb A \times\_ $ is the endofunctor modelling the explicity branching with the set $\mathbb A$ of actions. Moreover, if $B\colon\set \to\set$ has a Kleisli lifting $\overline B$, then a $TB$-coalgebras living in $\set$ could also be viewed as a $\overline B$-coalgebra living in the Kleisli category $\klcat (T)$ and %\cite{JACOBS2004167} for trace equivalence and developed further in a greater generality by Hasuo et al.\thinspace \cite{HasuoEtal:GenTraceSemantics}. Moreover, it was shown, in the context of labelled transition systems, that
the unique coalgebra homomorphism, in the context of LTSs, maps a state to the set of traces generated by the state \cite{JACOBS2004167}. This idea was developed further  by Hasuo et al.\thinspace \cite{HasuoEtal:GenTraceSemantics} by showing the behavioural equivalence in the chosen Kleisli category coincides with (probabilistic) language equivalence; in other words, language equivalence for various types of automata can be captured coinductively.

In a similar spirit,  one can recover various forms of decorated trace equivalences coinductively by moving from the category $\set$ of sets to an Eilenberg-Moore category (cf.\  \cite{BonchiEtal2012:finalSemDecoratedTraces,JacobsEtal2015:trace-det}) and graded algebras (cf.\  \cite{mps:trace-semantics-graded-monads,DorschEA19}) induced by graded monads. Nevertheless, these coinductive characterisations of decorated trace equivalences based on Eilenberg-Moore categories and graded monads \cite{graded-behEqGames} require a preprocessing step to determinise the given coalgebra (generalising the well-known determinisation procedure of an automaton); as a result, there is an exponential blowup on the underlying state space unlike in the Kleisli case.

In this paper, our aim is to use a coalgebraic machinery to synthesise decorated trace semantics (language equivalence, failure and ready equivalences) for conditional transition systems (CTS).
CTSs \cite{CTS-original,CTS_withupgrades-TASE17,CTSwithUpgrades2020,CTS-coalgebra-2018} are generalisations of traditional LTSs where each transition is guarded by a condition. CTSs come in two flavours based on whether the set of conditions are ordered or unordered. In the unordered case, CTSs and featured transition systems \cite{FTS2013}, a well-known operational model for software product lines, are equally expressive and we are able to characterise the above three decorated trace equivalence coinductively. In the ordered case, CTSs can model adaptive software product lines where certain features (encoded as conditions) can get upgraded to better versions modelled by the order relation; in this case, we present coinductive characterisations of language and ready equivalences (but not for failure equivalences). Nonetheless, for both types of systems, decorated trace semantics coarser than conditional bisimilarity \cite{CTS-original,CTS_withupgrades-TASE17,CTSwithUpgrades2020,CTS-coalgebra-2018} are not yet developed.

%\todo{correct this paragraph}
Our first contribution is to model CTSs without upgrades as $T^G B$-coalgebras, where $B=\mathbb{A}\times\_ + O$ is the endofunctor modelling the explicit branching with the set $\mathbb{A}$ of actions/alphabet and the set $O$ of observations attributed to various notions of decorated traces. The essential difference with the case of an LTS is to model the implicit branching by a relative monad $T^G$ \cite{AltenkirchEtal:RelativeMonads} induced by the powerset monad and the writer comonad $G = \condset \times\_$. This is to handle the `conditional' transition relation $\rightarrow \subseteq X \times \mathbb A \times \condset \times X$ of a CTS with a set $X$ of state space. Operationally, a CTS (without upgrade) executes by selecting a condition and, henceforth, behaviour evolves like in an LTS. This state-transition structure enriched with conditions from $\condset$ can be modelled as a coalgebra of type $\condset\times X \to \power(\condset \times (\mathbb A \times X + O))$; or simply as an arrow $X \to \mathbb A\times X + O$ in the Kleisli category $\klcat (T^G)$ induced by the relative monad $T^G$. The set $O$ of observations attributed to various notions of decorated traces (cf.\thinspace Section~\ref{sec:casestudy-cts}). %from a fixed set of conditions $\condset$ in a CTS ${\rightarrow} \subseteq X \times \mathbb{A} \times \condset \times X$.
%Inspired by \cite{HasuoEtal:GenTraceSemantics}, we resolve this by taking $T^G$ to be the relative monad \cite{AltenkirchEtal:RelativeMonads} induced by the powerset monad and the writer comonad $G = \condset \times\_$---so
Now the final coalgebra homomorphism $X \to \fseq {\mathbb A} \times O $ in $\klcat (T^G)$ is a function of type
$\condset \times X \to \power (\condset\times \fseq {\mathbb{A}} \times O)$
matching our intuition of mapping a state $x\in X$ and a condition to the set of decorated traces generated by $x$ and the condition obtained after executing a decorated trace (cf.\thinspace Theorem~\ref{thm:cts-noupgrades-coinductive}). %The relative monad $T^G$ models implicit branching (side effects) and $B=\mathbb{A}\times\_ + O$ is the endofunctor modelling the explicit branching with the set $\mathbb{A}$ of actions/alphabet and the set $O$ of observations attributed to various notions of decorated traces.

For CTSs with upgrades we move to the category $\pos$ of partially ordered sets and order preserving functions as morphisms with $T$ fixed to be the downset monad (cf.\thinspace \cref{sec:casestudy-cts}). However, the difference with \cite{CTS-coalgebra-2018} %\textit{op.\thinspace cit.}
is that we consider relative monads in this paper and thus, coalgebras in the Kleisli category $\klcat (\gmonad T)$ induced by a relative monad $\gmonad T$. Just like one needs to define a Kleisli lifting of an endofunctor in the classical case \cite{JACOBS2004167,HasuoEtal:GenTraceSemantics}, we prove a similar result (cf.\ Lemma~\ref{lem:KlLiftingRelMon}) in the context of relative monad $\gmonad T$. The conditions, though, of this lemma on the existence of Kleisli lifting $\widehat B \colon \klcat (\gmonad T) \to \klcat (\gmonad T)$ are quite strong; for instance, $G$ does not preserve $B=\mathbb{A}\times\_ + O$ because $GBX\not\cong BGX$ even when $O=1$.

Despite this hurdle we are able to reduce the problem of defining a Kleisli lifting $\widehat B$ for the  endofunctor $B=\mathbb{A}\times\_ + O$ in the context of a relative monad to the classical notion of Kleisli lifting \cite{HasuoEtal:GenTraceSemantics,JACOBS2004167,Mulry1994:Kleisli}. Furthermore, we were able to use a result by Freyd \cite{Freyd_1992} to prove that the initial algebra and final coalgebra for the functor $\widehat B$ coincide under the conditions that $\klcat (T)$ is $\cppo$-enriched and the $\omega$-directed joins commute with coproducts (cf.\  \cref{thm:finalcoal-machine}). These form the second contribution of the paper---paving a way to characterise decorated trace equivalences in a coinductive manner.

Our final contribution is to provide a recipe to construct a Kleisli lifting of a functor $F\colon\cat C \to \cat C$ in the context of just a monad $T\colon \cat C \to \cat C$. The key idea here is the following correspondence
% \vspace{-0.4cm} %final version
\[
\infer={X \times IY \to \Omega}{X \to TY}
\text,
% \vspace{-0.3cm} %final version
\]
which says that a Kleisli arrow is in one-to-one correspondence with a predicate over $X\times IY$
(here, $I\colon\cat C \to \cat C$ is an endofunctor and $\Omega$ is a truth value object in the category $\cat C$).
%Actually, we chose the relative Kleisli construction $ \klcat(T^G) $ instead of the ordinary Kleisli approach $ \klcat( T(\_)^\condset ) $ to make this approach work\todo{Do I remember this motivation correctly? DL}.
Once we have this correspondence, we can employ techniques from coalgebraic modal logic (like predicate and relation liftings) to define a distributive law $\vartheta$ of type $FT \Rightarrow TF$,  which is equivalent to a Kleisli lifting of $F$ (cf.\  \cite{Mulry1994:Kleisli}). In particular, under certain technical assumptions (\ref{assum:icat}-\ref{assum:plift}) we are able to define $\vartheta$ (cf.\  \cref{prop:dlaw-natural}) as the transpose (under the above correspondence) of a relation lifting applied to the relation $\inObj X \colon TX \times IX \to \Omega$. This relation coincides with the membership relation when $T=\power$ is the powerset monad and $I=\text{Id}$ is the identity functor on $\set$. Moreover, if the relation lifting preserves the diagonal relation $\Delta$ (defined as the transpose of the unit of $T$) and the relation composition (defined using the Kleisli composition), then the constructed natural transformation $\vartheta$ is well-behaved with the unit (cf.\  Lemma~\ref{lem:unit-wellbehaved}) and multiplication (cf.\  Lemma~\ref{lem:mult-wellbehaved}) of the monad $T$, respectively. As a result, in the context of $\cat C=\set$ or $\cat C=\pos$, we obtain that the constructed natural transformation $\vartheta$ is a distributive law of type $FT \Rightarrow TF$ whenever $F$ preserves the weak pullback squares.

\section{Preliminaries}\label{sec:prelim}
%\todo[inline]{Introduce coalgebraic preliminaries, Kleisli category and Kleisli lifting.}
The objective of this section is to set the notations for this paper and recall the preliminaries
related to coalgebraic modelling in a Kleisli category from \cite{HasuoEtal:GenTraceSemantics}.

We assume familiarity with basic category theory and the theory of coalgebras.  We use meta-predicates
$X, Y\in \cat C$ and $f\in\cat C(X,Y)$ to denote objects $X, Y$  and an arrow $f$ of the category $C$, respectively.
If $X, Y\in \cat C$ have a coproduct, we write $\iota_X\colon X \to X + Y$ for the inclusion map.
Dually, we write the projection map $\proj_X\colon X \times Y \to X$ whenever the product $X \times Y$ exists.
Moreover, when $f'\colon X' \to Y'\in \cat C$ and the coproducts $ X + X' $ and $ Y + Y' $ exist, we denote by $f+f'\colon X + X' \to Y + Y'$  the unique arrow from
the universal property of coproduct $X + X'$ such that the equations
$\iota_Y \circ f =(f+f') \circ\iota_X$ and  $ \iota_{Y'} \circ f' = (f+f') \circ \iota_{X'}$ hold.
Dually, if $ X \times X' $ and $ Y \times Y' $ exist, we define $
    f \times f'\colon  X \times X' \to Y \times Y'
$  to be the unique map given by $ f \circ \proj_X, f' \circ \proj_{X'} $ and the universal property of $Y \times Y'$.

\subsection{Coalgebras in a Kleisli category}\label{ssec:coalg_Kleisli}

We fix a category $\cat C$ and a monad $(T,\eta,\mu)$ on $\cat C$. Recall the Kleisli category $\klcat (T)$ induced by $T\colon \cat C \to \cat C$:
\[
\infer={X \in \klcat (T)}{ X \in \cat C} \qquad \infer={f\colon X\to Y \in \klcat (T)}{ f\colon X \to TY \in \cat C}.
\]
The Kleisli composition $g\bullet f$ of two arrows $f\colon X\to Y,g\colon Y \to Z$ is given by the composition $\mu_Z\circ Tg \circ f$.

Throughout this section, we fix a coalgebra $c\colon X\to TBX \in\cat C$, where $B$ is an endofunctor on $\cat C$. Typical examples are nondeterministic automata (NDAs), when $B=\mathbb{A} \times \_ + 1$ and $T =\power$, or
their probabilistic/weighted variant, when $T$ is sub-distribution monad \cite{HasuoEtal:GenTraceSemantics} or semiring monad \cite{JacobsEtal2015:trace-det}.

It is well known---for instance in the context of NDAs---that the coalgebra homomorphisms
correspond to functional bisimulations which are too strong to capture language equivalence
(either by taking the span or cospan of coalgebra homomorphisms). This mismatch is
avoided, as first noted in \cite{JACOBS2004167}, by moving to the
Kleisli category $\klcat (\power)$. In particular, NDAs can also be seen as the coalgebra
$X\to \overline BX \in \klcat (\power)$, where $\overline B$ is the  Kleisli lifting of
the endofunctor $B= \mathbb{A}\times \_ + 1$.

In general, a functor $\overline B \colon \klcat (T) \to \klcat (T)$ is a \emph{Kleisli lifting} of
$B \colon \cat C \to \cat C$ whenever the following square (drawn on the left) commutes, i.e.\thinspace $\overline B \circ L' = L' \circ B$.
Here $L'$ is the free functor that maps an object $X$ to the free algebra $(TX,\mu_X)$ and
is left adjoint to the forgetful functor $R' \colon \klcat (T) \to \cat C$ with $R'(X) = TX$ (for $X\in\klcat (T)$) and $R'(f) = \mu'_X \circ Tf $ (for $f\colon X \to Y \in \klcat (T)$).

\begin{tikzcd}
\klcat (T)  \arrow[r,"\overline B"] & \klcat (T)\\
\arrow[u,"L'"] \cat C \arrow[r,"B"] & \arrow[u,"L'"'] \cat C
\end{tikzcd}
\qquad\qquad
\begin{tikzcd}
TB \arrow[r,"\vartheta"] & BT\\
\arrow[u,"\eta_B"] \arrow[ru,"B\eta"']
B
\end{tikzcd}\eqnum\label{eq:unit-wellbehaved}
\qquad\qquad
\begin{tikzcd}
BTT \arrow[r,"\vartheta_T"] \arrow[d,"B\mu"'] & TBT \arrow[r,"T\vartheta"]& TTB
\arrow[d,"\mu_B"]\\
BT\arrow[rr,"\vartheta"] && TB
\end{tikzcd}\  \eqnum\label{eq:mult-wellbehaved}

Moreover, $\overline B$ is a Kleisli lifting of $B$ \cite{HasuoEtal:GenTraceSemantics,Mulry1994:Kleisli}
iff there is a natural transformation $\vartheta \colon BT \Rightarrow TB$ satisfying the
laws indicated above in the middle and on the right. Such natural transformations were coined
 \kllaws{} in \cite{JacobsEtal2015:trace-det}. The upshot of having a Kleisli lifting
is that (probabilistic) language equivalence can be captured in a coinductive manner
\cite{HasuoEtal:GenTraceSemantics}, i.e.\ whenever the final $\overline B$-coalgebra exists.

%In \textit{op.\thinspace cit.}, the author\todo{is that Harsh or the author of one of the references? In the former case rather "one of the authors"} 
Hasuo et al.\thinspace \cite{HasuoEtal:GenTraceSemantics} presented two conditions (of increasing strengths) on a Kleisli lifting $\overline B$ that ensured when the initial $B$-algebra in $\cat C$ (if it exist)
coincides with the final $\overline B$-coalgebra in $\klcat (T)$. In this paper, we will use
the following result due to Freyd \cite{Freyd_1992} and generalise it to the level of
relative monads. The other condition in \cite[Theorem~3.3]{HasuoEtal:GenTraceSemantics} requires that $\overline B$ is locally monotone instead of locally continuous; though we work with stronger assumption since  $B$ in our case studies will be locally continuous.% (just like in the case of \cite{HasuoEtal:GenTraceSemantics}).

\begin{definition}
A category $\cat C$ is a $\cppo$-enriched category whenever its hom-set forms a $\omega$-cpo
with a bottom and the composition of arrows is a continuous function. In particular,
\begin{itemize}
\item for each $X,Y\in\cat C$, the set $\cat C(X,Y)$ is partially ordered $\preceq_{X,Y}$
with a bottom $\bot_{X,Y}\in \cat C(X,Y)$ (we drop the subscripts whenever it is clear from
the context);
\item for every increasing $\omega$-chains $(f_i\in\cat C(X,Y))_{i\in\nat}$ (i.e.\
$f_i \preceq f_{i+1}$), the join $\bigvee_{i\in\nat} f_i\in\cat C(X,Y)$ exists.
\item
    for every increasing $\omega$-chains $(f_i\in\cat C(X,Y))_{i\in\nat}$ and every
    $g\in\cat C(Y,Y'),h\in\cat (X',X)$ we have
    $g\circ (\bigvee\nolimits_{i\in\nat} f_i) =\bigvee\nolimits_{i\in \nat}  g\circ f_i \quad \text{and}\ \quad
    (\bigvee\nolimits_{i\in\nat} f_i) \circ h = \bigvee\nolimits_{i\in\nat} f_i\circ h$.
\end{itemize}
\end{definition}
\begin{theorem}[\cite{Freyd_1992,HasuoEtal:GenTraceSemantics}]\label{thm:cppo-klcat}
Let $\klcat (T)$ be a $\cppo$-enriched category whose composition is left strict (i.e.\ $\bot_{Y,Y} \bullet f=\bot_{Y,Y}$ for every $f\in\klcat (T)(X,Y)$) and
$\overline B\colon \klcat (T) \to \klcat (T)$ be a locally continuous endofunctor. Then an
initial algebra $\alpha\colon B(\mu_B) \xrightarrow{\cong} \mu_B \in \cat C$ (if it exists) induces
a final coalgebra $B(\alpha) \colon \mu_B \to \overline B (\mu_B) \in \klcat (T)$.
\end{theorem}

%\todo[inline]{Recall the main result of \cite{HasuoEtal:GenTraceSemantics} and instantiate it
%for the behaviour functor $B=\mathbb{A}\times X + C$ where $\mathbb{A},C$ are some fixed sets.}

%\todo[inline]{Apply the above result in characterising failure/ready equivalences (I think these
%characterisations via Kleisli category are new; note use $C=\power A$).}
\subsection{Decorated trace equivalences coinductively}
In this subsection, we apply \cref{thm:cppo-klcat} to characterise failure and ready
equivalences using coinduction, i.e.\ we will characterise these equivalences as the equivalence induced by a unique coalgebra homomorphism from the underlying coalgebra to the final coalgebra. Though the  presentation is new and the results follow directly from the above theorem, but we do \emph{not} claim novelty (perhaps this is folklore). To the best of our knowledge, these decorated trace equivalences were only
characterised by considering coalgebras in Eilenberg-Moore categories
\cite{BonchiEtal2012:finalSemDecoratedTraces,JacobsEtal2015:trace-det} or in the setting of graded algebras \cite{DorschEA19}, but not in a Kleisli setting.

Throughout this subsection, we fix the endofunctor $B=\mathbb{A}\times\_ \ +\ O$ where $\mathbb{A}$ and $O$ are some fixed sets with $O$ indicating
the observations that make these decorated trace equivalences distinct among themselves. So a
labelled transition system (LTS) enriched with observations from $O$ is a coalgebra
$c\colon X \to \power B X$.

\begin{proposition}\label{prop:IAlgExist-Set}
For the endofunctor $B=\mathbb{A} \times \_ \ + \ O$ on $\set$, the initial algebra exists and is given
by $\mu_B = \fseq{\mathbb{A}} \times O$ (the product of the sets of finite words induced by $\mathbb{A}$ and
observations). Moreover, the algebra $h+h'\colon B\mu_B \xrightarrow{\cong} \mu_B$
is given by $h'(o) = (\epsilon, o)$ and $h(a,w,o)=(aw,o)$ (for $a\in \mathbb{A},w\in\fseq{\mathbb{A}}, o\in O$).
\end{proposition}
Moreover, $\klcat (\power)$ is a $\cppo$-enriched category (cf.\
\cite{HasuoEtal:GenTraceSemantics}) where the order is given by the subset inclusion and
$\bot$ is given by the empty relation (recall that $\klcat (P)$ is isomorphic to the category
$\cat{Rel}$ of sets as objects and relations as morphisms). The functor $B$ has a Kleisli lifting $\overline B$, which acts on a relation $f\colon\mathbf {Rel}(X,Y)$ as
follows (which can be derived using the machinery developed in \cref{sec:Kl-lift}; see Example~\ref{ex:machine-lift-abstractly}):
\begin{equation}\label{eq:machine-lifting}
\overline Bf = \{(o,o) \mid o\in O\} \cup \{((a,x),(a,y)) \mid a\in \mathbb{A}\land x\mathrel f y\}.
\end{equation}
Now \cref{thm:cppo-klcat} becomes applicable and we have
\begin{proposition}
The final coalgebra for $\overline B$ exists in $\klcat (\power)$ and is given by
$\fseq{\mathbb{A}} \times O$.
\end{proposition}
Moreover, it is instructive to verify that the unique coalgebra homomorphism $f\colon X \to \fseq{\mathbb{A}} \times O \in \klcat(\power)$ from
the coalgebra $c\colon X \to \overline B X \in\klcat (\power)$ maps a state $x$ to a set
of tuples $(w,o)\in \fseq{\mathbb{A}} \times O$ such that $o$ is an observation after performing the
trace $w$ from the state $x$. In addition, by modelling
\emph{refusal sets} \cite{g:linear-branching-time} in the coalgebra map $c$ when $O=\power {\mathbb{A}}$, i.e.\
for any $x \in X$ we impose on $c(x) \subseteq \mathbb{A} \times X + \power\mathbb A$ the condition
\[
    o \in c(x) \iff o \ \text{is the set of actions disallowed from the state}\ x
\]
for any $o \in O$, we obtain that the unique coalgebra homomorphism $f$ maps a state to its failure
pairs \cite{g:linear-branching-time}. A \emph{failure pair} of a state $x$ is a tuple $(w,o)$ where $w$ is a trace starting from $x$ to some $x'$ and, moreover, $o$ is a `refusal' set of actions disallowed from $x'$. As a result, we get a coinductive characterisation
of failure equivalence. Similarly, modifying the above coalgebra map to a map with the condition
\[
            o \in c(x)
    \iff    o \ \text{is the set of actions enabled from the state}\ x
\]
for any $o \in O$, obtain that the unique coalgebra homomorphism $f$ maps a state to its ready pairs \cite{g:linear-branching-time}. A \emph{ready pair} of a state $x$ is a tuple $(w,o)$ where $w$ is a trace starting from $x$ to $x'$ and $o$ is a `ready' set of actions enabled from the state $x'$. Thus, obtaining a coinductive characterisation of ready equivalence.
\section{Relative monads, Kleisli categories, and Kleisli liftings}\label{sec:RelMon}
In \cref{sec:casestudy-cts} it will become apparent that CTSs are modelled as coalgebras living in a
Kleisli category induced by a relative monad. Relative monads in Computer Science were
introduced in \cite{AltenkirchEtal:RelativeMonads}; in particular, they worked out the so-called
Kleisli and Eilenberg-Moore constructions of a relative monad. Nevertheless, the question of
Kleisli lifting of an endofunctor was not considered in \textit{op.\thinspace cit.} and is particularly relevant for
coalgebraic modelling of CTSs with and without upgrades. Therefore, in this section, we are
going to recall the Kleisli construction of a relative monad from \cite{AltenkirchEtal:RelativeMonads}
and give sufficient conditions that ensure that the resulting functor is a Kleisli lifting of a given endofunctor.

\begin{definition}
  Given a functor\footnote{Our presentation of relative monad is an instance of a more general formulation in \cite{AltenkirchEtal:RelativeMonads} where $G$ is not necessarily an
endofunctor.} $G\colon \cat C \to \cat C$, then a \emph{$G$-relative monad} \cite{AltenkirchEtal:RelativeMonads} is given by the following data:
  \begin{enumerate}
    \item
        an object mapping $T\colon \cat C \to \cat C$;
    \item
        for every object $X\in \cat C$, there is a \emph{unit} map $\eta_X \in \cat C(GX, TX)$;
    \item
        for every arrow $f\in \cat C (GX,TY)$ there is a map $f^\sharp\in\cat C(TX,TY)$ called the \emph{Kleisli lifting} of $f$ satisfying the unit and associative laws, i.e.\ for any $g\in\cat C(GY, TZ)$ the following
        diagrams commute:
  % \vspace{-1em}
  \begin{subequations}
  \noindent
  \hspace{-4em}
  \begin{tabularx}{\textwidth}{@{}XXX@{}}
    \begin{equation}\vphantom{\begin{tikzcd}[ampersand replacement=\&]
        TX\arrow[d,"f^\sharp"] \arrow[r,"(g^\sharp\circ f)^\sharp"] \& TZ\\
        TY\arrow[ru,"g^\sharp"']
    \end{tikzcd}}\begin{tikzcd}[ampersand replacement=\&]
            \label{tikz:leftUnit}
        GX \arrow[r,"\eta_X"] \arrow[d,"f"']\& \arrow[ld,"f^\sharp"] TX \\
        TY \&
    \end{tikzcd}\end{equation}
    &
    \begin{equation}\vphantom{\begin{tikzcd}[ampersand replacement=\&]
        TX\arrow[d,"f^\sharp"] \arrow[r,"(g^\sharp\circ f)^\sharp"] \& TZ\\
        TY\arrow[ru,"g^\sharp"']
    \end{tikzcd}}\begin{tikzcd}[ampersand replacement=\&]
            \label{tikz:rightUnit}
       \& TX \arrow[d,shift left=1ex, "\eta_X^\sharp"]
        \arrow[d,shift right=1ex, "\id{TX}"'] \\
        \& TX
    \end{tikzcd}\end{equation}
    &
    \begin{equation}\begin{tikzcd}[ampersand replacement=\&]
            \label{tikz:associativity}
        TX\arrow[d,"f^\sharp"] \arrow[r,"(g^\sharp\circ f)^\sharp"] \& TZ\\
        TY\arrow[ru,"g^\sharp"']
    \end{tikzcd}\end{equation}
  \end{tabularx}
  \end{subequations}
  \end{enumerate}
\end{definition}

Just like how a traditional monad gives rise to a Kleisli category, so does the relative monad in the manner explained next. Given a $G$-relative monad $T$, its \emph{Kleisli category}, denoted $\klcat (\gmonad T)$, is given by objects from $\cat C$ and maps between any $X$ and $Y$ by maps between $GX$ and $TY$ in $\cat C$, i.e.\
\begin{equation}
        \label{eq:KleisliCat}
    \infer={X \in \klcat (\gmonad T)}{X \in \cat C} \qquad \infer={f\colon X \to Y \in \klcat (\gmonad T)}{f\colon GX \to TY \in \cat C}
\text.
\end{equation}
The identity morphism on $X$ is provided by $\eta_X$, which is a left and right unit according to \cref{tikz:leftUnit,tikz:rightUnit}, respectively.
Composition of two morphisms $X \xrightarrow{f} Y \xrightarrow{g} Z$ in $\klcat(\gmonad T)$ is given by $g^\sharp  \circ f$ in $\cat C$. Associativity of this composition is ensured by
\cref{tikz:associativity}.

The usual free-forgetful adjunction between the Kleisli category and its underlying base category gets a bit subtle in the presence of relative monads.
The notion of adjunction generalises to that of a \textbf{(left) $G$-relative adjunction} $ L \dashv R\colon G \to \cat D $ between an endofunctor $G$ on $\cat C$ and another category
$\cat D$.
It consists of two functors $L\colon \cat C \to \cat D$ and $R\colon \cat D \to \cat C$ such that we have the natural bijection
\[
    \infer={LX \to Y \in \cat D}{GX \to RY \in \cat C}
\text.
\]

\begin{proposition}
    For any endofunctor $G$ and monad $T$, both on $\cat C$,
  there is a $G$-relative adjunction $L\dashv R\colon G \to \klcat (\gmonad T)$, called the \textbf{relative Kleisli adjunction}, given by the functors $L,R$:
  \begin{subequations}
  \begin{align}
        \label{eq:relKleisli_embedding}
      LX &= X,    &  Lf &= \eta_{GY} \circ Gf,  \\
      RX &= TX,   & Rg &= g^\sharp,
  \end{align}
  \end{subequations}
  where $f\colon X\to Y$ and $g\colon GX \to TY$.
\end{proposition}
%
%\begin{proof}
%    The natural bijection of this adjunction is simply given by \cref{eq:KleisliCat}.
%\end{proof}
\subsection*{Every monad induces a relative monad}
Now fix an endofunctor $G$ on a category $\cat C$, then every monad $(T,\eta,\mu)$ on $\cat C$ gives rise to a relative monad $\gmonad T$ with $\gmonad T X = TG X$ (for
$X\in \cat C$), the unit given by $\eta_G$, and $f^\sharp =\mu_{GY}\circ Tf$ (for $f\colon GX \to TGY$). %All the relative monads in this paper are constructed this way.

\begin{proposition}{\cite[\S~2.4]{AltenkirchEtal:RelativeMonads}}
The three categories are formally related as follows:
    \[
\begin{tikzcd}
  \klcat(\gmonad T)\arrow[rd,"R"'] \arrow[r,"D"]&
  \klcat (T)\arrow[d,xshift=1.5ex,"R'"] \\
  \arrow[u,"L"]\cat C\arrow[r,"G"']& \arrow[u,xshift=-1.5ex,"L'","\dashv"']\cat C
\end{tikzcd}
\]
Where $L'\dashv R'$ is the classical Kleisli adjunction, $DX=GX$, and $Df=f$ (for $X,f\in\klcat (\gmonad T)$).% on objects and the identity on morphisms.
\end{proposition}
%\begin{proof}
%  \cite[\S~2.4]{AltenkirchEtal:RelativeMonads}
%\end{proof}

Note that the construction $\gmonad T$ resembles the ``Kleisli-like'' construction $K^T_{G, T}$ from Hirsch's thesis \cite[p.\ 44]{Hirsch2019:semantics_secureSoftware}, where it was assumed
that $G$ is a comonad.
He explores different ways to combine monads and comonads in programming language semantics and applies his results to security policies.

\subsection*{Kleisli lifting of an endofunctor}
In the sequel, all our relative monads are induced by monads; so, in this section, we explore how to extend a given endofunctor $B \colon \cat C \to \cat C$ to an endofunctor $\widetilde
B \colon \klcat (\gmonad{T}) \to \klcat (\gmonad{T})$. Just like in the traditional case, we say an endofunctor $\widetilde B \colon \klcat (\gmonad{T}) \to \klcat (\gmonad{T})$ is a
\emph{Kleisli lifting} of $B$ iff $\widetilde{B} \circ L = L \circ B$.

%\begin{wrapfigure}[5]{R}{0.3\textwidth}
%\vspace{-0.8cm}
\begin{equation}\label{eq:def-Btilde}
\begin{tikzcd}[ampersand replacement=\&]
GBX\arrow[d,"\rho_X","\cong"'] \arrow[r,dashed,"\widetilde{B}f"] \& TGBY\\
BGX \arrow[r,"\bar Bf"] \& TBGY \arrow[u,"T\inv{\rho}_Y","\cong"']
\end{tikzcd}
\end{equation}
%\end{wrapfigure}
\begin{lemmarep}\label{lem:KlLiftingRelMon}
  If $G$ preserves $B$, i.e.\ there is a natural isomorphism $\rho\colon GB \cong BG$, the existence of a Kleisli lifting $\klcat (T) \xrightarrow {\overline B} \klcat(T)$ of $B$ implies the
existence of a Kleisli lifting $\klcat(\gmonad{T}) \xrightarrow{\widetilde{B}} \klcat(\gmonad{T})$ of $B$. In particular, $\widetilde BX = BX$ (for $X\in\cat C$) and $\widetilde Bf$ (for an arrow $f\colon X \to Y \in\klcat (\gmonad T)$) is defined as in \eqref{eq:def-Btilde}
Moreover, if $\overline B$ is locally continuous (when $\klcat (T)$ is $\cppo$-enriched), then so is $\widetilde B$ as defined above.%the Kleisli lifting $\widetilde B$ defined in the proof.
\end{lemmarep}
\begin{proof}
  %Let $\theta\colon BT \Rightarrow TB$ be the distributive law induced by the Kleisli lifting $\overline{B}$.
  %Consider the mapping $\widetilde{B} X = BX$ and the following diagram on the left
  Identity morphisms are preserved by $\widetilde B$, i.e. $\widetilde B = \eta_{GBX}$, as becomes apparent from the solid square
  \[
  \begin{tikzcd}
    %GBX\arrow[d,"\rho_X","\cong"'] \arrow[r,dashed,"\widetilde{B}f"] & TGBY &
    GBX\arrow[d,"\rho_X","\cong"'] \arrow[r,"\eta_{GBX}"'] \rar["\widetilde B (\id X)" near start, bend left, dashed] &[2em] TGBX\\
    %BGX \arrow[r,"\bar Bf"]& TBGY \arrow[u,"T\inv{\rho}_Y","\cong"'] &
    BGX \arrow[r,"\eta_{BGX}"]& TBGX \arrow[u,"T\inv{\rho}_X","\cong"']
  \end{tikzcd}
  \]
  (i.e.\ naturality of $\eta$), that $\rho_X$ is an isomorphism, and $\bar B\eta_{GX}=
\eta_{BGX}$.

 For the preservation of the composition of arrows, consider the following diagram with $f\colon X\to Y,g\colon Y\to Z \in \klcat (\gmonad{T})$.
  \[
  \begin{tikzcd}[execute at end picture={\node[] at ($(A.south) + (2.25cm,-0.6cm)$) {$\overline{B}g\bullet\overline{B}f=\overline B(g\bullet f)$};}]
    GBX\arrow[d,"\rho_X","\cong"'] \arrow[r,"\widetilde{B}f"] & TGBY \arrow[r,equal] & TGBY \arrow[d,"T\rho_Y"]\\
    BGX \arrow[r,"\bar Bf"]
    \arrow[rrrr,rounded corners,
    to path ={|- ([yshift=-2ex]\tikztostart.south) -- ([yshift=-2ex]\tikztotarget.south) -| (\tikztotarget.south)},
    at end]
    & |[alias=A]| TBGY \arrow[u,"T\inv{\rho}_Y","\cong"'] \arrow[r,equal] & TBGY \arrow[r,"T\overline{B}g"] & TTBGZ \arrow[r,"\mu_{BGZ}"]& TBGZ
  \end{tikzcd}
  \]
  Its commutativity follows directly from the functoriality of $\overline{B}$.

  Now suppose $\klcat (T)$ is $\cppo$-enriched and $\overline B$ is locally continuous. Note that $\klcat (\gmonad T)$ is also $\cppo$-enriched since $\klcat (\gmonad T) (X,Y) = \klcat (T)(GX,GY)$ and all the order-theoretic structures on the former is inherited from the latter. Moreover,
  \[
  \bigvee_{i\in \nat} \widetilde B(f_i) = \bigvee_{i\in \nat} T\inv{\rho}_Y \circ \overline Bf_i \circ \rho_X = T\inv{\rho}_Y \circ \left(\bigvee_{i\in\nat }\overline Bf_i \right) \circ \rho_X = \widetilde B \left(\bigvee_{i\in\nat}f_i\right).
  \]
\end{proof}
Unfortunately, unlike in the traditional case, we have to enforce some restriction on $G$ (cf.\  Lemma~\ref{lem:KlLiftingRelMon}) to characterise a Kleisli lifting in terms of
certain distributive laws. Note that this condition is, perhaps, not that surprising when compared to other existing results that lift traditional results known on monads/adjunctions to
relative monads/adjunctions. For instance, the well known result that colimits are preserved by left adjoints does \emph{not} hold in general in the context of relative adjunctions; it is only those
colimits that are preserved by $G$ are preserved by the left adjoint in this new setting \cite{ULMER1968:DenseAndRelAdjoint}.

\subsection*{Kleisli lifting of Machine endofunctor $B=\mathbb{A}\times\_ + O$}
The condition in Lemma~\ref{lem:KlLiftingRelMon} that $G$ preserves the endofunctor $B$ is too strong because $GB X \not\cong BGX$ (for $X\in\set$, any nonempty set $O$, and $G = \condset \times \_ $ with $\condset \neq \emptyset$). In this section, we further impose the restriction on $G$ to preserve coproducts, which allows us to define a Kleisli lifting for the machine endofunctor. Assume that the working category $\cat C$ has binary products and coproducts, so that we can define the
machine endofunctor $B = \mathbb{A}\times\_ + O$, where $\mathbb{A},O\in\cat C$ are two fixed objects in the category $\cat C$. For brevity, the functor $ \mathbb{A}\times\_ $ is denoted by $A$.

Throughout this section, we further assume that
\begin{enumerate}[label=\textbf{A\arabic*}]
  \item\label{assum:GpresCoproducts} the functor $G$ preserves coproducts;
  \item\label{assum:GpresA} the functor $G$ preserves $A$ (cf.\  Lemma~\ref{lem:KlLiftingRelMon});
  \item\label{assum:barAExists} $\overline A\colon\klcat (T) \to \klcat(T)$ is a Kleisli lifting of $A$ (thus $\widetilde{A}\colon\klcat(\gmonad{T}) \to \klcat(\gmonad{T})$ exists).
\end{enumerate}
These assumptions allow us to define the mapping $\widehat{B}$, which will become our lifting: It maps an object $X$ to $BX$ and for a given arrow $f\colon X\to Y \in \klcat(\gmonad{T})$ we
define (note \ref{assum:GpresA} ensures that Lemma~\ref{lem:KlLiftingRelMon} is applicable for the endofunctor $A$):
\begin{equation}\label{eq:MachineEndo_def}
    %\begin{tikzcd}%[column sep=huge]
    %  G(AX + C) \arrow[d,equal] \arrow[rrr,"\widehat{B}f"]&&& TG(AY+C)\\
    %  GAX + GC \arrow[r,"\rho_X+\eta_{GC}"']& AGX + TGC \arrow[r,"\overline Af+TGC"']& TAGY + TGC \arrow[r,"T\inv\rho_{Y}+TGC"']&
    %  \arrow[u,hookrightarrow,"TG\iota_{AY}+TG\iota_{C}"]TGAY
    %  +TGC
    %\end{tikzcd}
    \begin{tikzcd}[column sep=huge]
      G(AX + O) \arrow[r,"\widehat{B}f"]& TG(AY+O)\\
      GAX + GO
        \arrow[r,"\widetilde{A}f + \eta_{GO}"']
        \uar["G\iota_{AX} \coDiagOp G\iota_O", "\cong"']
      & \arrow[u,hookrightarrow,"TG\iota_{AY} \coDiagOp  TG\iota_{O}"]TGAY +TGO
    \end{tikzcd}
\end{equation}
where $f \coDiagOp  f'\colon X + X' \to Y$, ``codiagonal'', is defined as the universal arrow of two morphisms $f\colon X \to Y$ and $f'\colon X' \to Y$ with joined codomain. It is actually the map given from the pair $(f, f')$ as the adjunct of the adjunction between the coproduct and the diagonal functor $X \mapsto (X, X) $ from $\cat C$ to $\cat{Cat}( \{1,2\}, \cat C)$.
Later in \cref{sec:Kl-lift} we will also use the dual version $\diagOp$ of $\coDiagOp$.

\begin{theoremrep}\label{thm:KlExtMachineEndo}
  The above mapping $\widehat B\colon\klcat(\gmonad{T}) \to \klcat(\gmonad{T})$ is a functor. Moreover, $\widehat B$ is also a Kleisli lifting of $B$, i.e.\ $\widehat{B}\circ L = L \circ
B$.
\end{theoremrep}
\begin{proof}
  For preservation of identity observe for any object $X$ in $\cat C$ that by uniqueness of coproducts
  \begin{align*}
            \widehat B (\id{X}) \qquad
        &\refrel{eq:MachineEndo_def}=
            (TG\iota_{AY} \coDiagOp  TG\iota_O) \circ ( \widetilde A (\id{X}) + \eta_{GO} )
    \\  &=  TG\iota_{AY} \circ \widetilde A (\id{X}) \coDiagOp  TG\iota_O \circ \eta_{GO}
    \\  &\stackrel{\mathllap{\text{\textbf{A1}}}}=
            T\iota_{GAY} \circ \widetilde A (\id{X}) \coDiagOp  T\iota_{GO} \circ \eta_{GO}
    \\  &=  T\iota_{GAY} \circ \eta_{GAX} \coDiagOp  T\iota_{GO} \circ \eta_{GO}
    \text.
  \end{align*}
  Since $\eta$ is a natural transformation, the following diagram without the dashed arrow
  \begin{equation}\label{eq:coproduct_unit}
  \begin{tikzcd}
        TGAX \ar[r, "T\iota_{GAX}"]
      & T(GAX + GO)
      &[4em] TGO \ar[l, "T\iota_{GO}"']
    \\
        GAX \ar[r, "\iota_{GAX}"'] \ar[u, "\eta_{GAX}"]
      & GAX + GO
        \ar[u, "\eta_{GAX + GO}", bend left]
        \ar[u, "T\iota_{GAX}\circ \eta_{GAO} \coDiagOp  T\iota_{GO} \circ \eta_{GAX}" right, bend right, dashed]
      & GO \ar[l, "\iota_{GO}"] \ar[u, "\eta_{GAX}" right]
  \end{tikzcd}
  \end{equation}
  commutes.
  The diagram commutes as well when the middle arrow is replaced by the dashed one.
  As the latter is unique by the universality of coproducts, it equals the former one. Hence $ \widehat B (\id{X}) = \eta_{GAX + GO} $, which is the identity arrow in $\klcat(\gmonad T)$.

  For the preservation of composition, first, we observe in parallel to the diagram with the unit $\eta$ above that, since the multiplication $\mu$ is a natural transformation, for any object $Z$ in $\cat C$ the non-dashed arrows in the diagram
  \begin{equation*}
      \begin{tikzcd}[column sep=huge]
            &   TTGAZ + TTGO
                \ar[d, "TT\iota_{GAZ} \coDiagOp  TT\iota_{GO}" below left, dashed]
                \ar[dd, "\substack{TG\iota_{AZ} \circ \mu_{GAZ} \\ {} \coDiagOp  TG\iota_{O} \circ \mu_{GO}}" above right, bend left, dashed] &
        \\[2em]
                TTGAZ    \ar[d, "\mu_{GAZ}" left]
                    \ar[ru, "\iota_{TTGAZ}", bend left, dashed] \ar[r, "TT \iota_{GAZ}" below]
            &   TT(GAZ + GO)\ar[d, "\mu_{G(AZ + O)}" left]
            &   TTGO \ar[d, "\mu_{GO}" left]
                    \ar[lu, "\iota_{TTGO}"', bend right, dashed] \ar[l, "TT \iota_{GO}"]
        \\
                TGAZ \ar[r, "T \iota_{GAZ}" below]
            &   TG(AZ + O)
            &   TGO \ar[l, "T \iota_{GO}"]
      \end{tikzcd}
  \end{equation*}
  commute.
  By the uniqueness of $TG\iota_{AZ} \circ \mu_{GAZ} \coDiagOp  TG\iota_{O} \circ \mu_{GO}$ and the universal property of $TT\iota_{GAZ} + TT\iota_{GO}$, we have
  \begin{equation}\label{eq:coproduct_mult}
      (TG\iota_{AZ} \circ \mu_{GAZ}) \coDiagOp  (TG\iota_{O} \circ \mu_{GO}) = \mu_{G(AZ + O)} \circ (TT\iota_{GAZ} \coDiagOp  TT\iota_{GO})
    \text.
  \end{equation}
    \begin{subequations}
  Further for any morphisms $h\colon V \to W$ and $h'\colon V' \to W$ in $\cat C$ with joined codomain consider the diagram
  \[
    \begin{tikzcd}[column sep=huge]
        & TV \ar[ld, "\iota_{TV}"', bend right] \ar[d, "\iota_{V}" near end] \ar[dr, "Th", bend left] &
    \\[1em]
            TV + TV' \ar[r, "T\iota_{V} \coDiagOp  T\iota_{V'}"', dashed] \ar[rr, "Th \coDiagOp  Th'"' near start, dashed, bend left]
        &   T(V+ V')
                \ar[r, "T(h \coDiagOp  h')" below]
        &   TW
    \\[1em]
        & TV' \ar[lu, "\iota_{TV'}", bend left] \ar[u, "\iota_{V'}"]  \ar[ur, "Th'"', bend right] &
    \end{tikzcd}
  \]
  where the dashed arrows are given by the universal property of coproducts.
  By uniqueness of such arrows, we have
  \begin{equation}\label{eq:functor_coproduct}
            T(h \coDiagOp  h') \circ (T\iota_{V} \coDiagOp  T\iota_{V'})
      =     Th \coDiagOp  Th'
  \end{equation}
  (just as $T$ is a functor, not using its monadic properties).
  Thus for $h + h'$ of two morphisms $h\colon V \to W$ and $h'\colon V' \to W'$ we get
  \begin{equation}  \label{eq:functor_coDiagOp}
      T(h + h') \circ (T\iota_{V} \coDiagOp  T\iota_{V'}) = T\iota_{W} \circ Th \coDiagOp  T\iota_{W'} \circ Th'
    \text.
  \end{equation}
  \end{subequations}
  Now we infer preservation of composition by verifying that for any two morphisms $X \xrightarrow{f} Y \xrightarrow{g} Z$
  \begin{align*}
            \widehat B (g) \bullet \widehat B (f)
        &=  \widehat B (g)^\sharp \bullet \widehat B (f)
    \\  &\refrel{eq:MachineEndo_def}=
                ( (TG\iota_{AZ}\coDiagOp TG\iota_{O}) \circ (\widetilde{A}g + \eta_{GO}) )^\sharp
            \circ (TG\iota_{AY}\coDiagOp TG\iota_{O}) \circ (\widetilde{A}f + \eta_{GO})
    \\  &\stackrel{\mathllap{\text{\textbf{A1}}}}=
                ( (T\iota_{GAZ}\coDiagOp T\iota_{GO}) \circ (\widetilde{A}g + \eta_{GO}) )^\sharp
            \circ (T\iota_{GAY}\coDiagOp T\iota_{GO}) \circ (\widetilde{A}f + \eta_{GO})
    \\  &=      \mu_{G(AZ + O)}
            \circ T\big( (T\iota_{GAZ}\coDiagOp T\iota_{GO}) \circ (\widetilde{A}g + \eta_{GO}) \big)
            \circ (T\iota_{GAY}\coDiagOp T\iota_{GO}) \circ (\widetilde{A}f + \eta_{GO})
    \\  &=      \mu_{G(AZ + O)}\circ T(T\iota_{GAZ}\coDiagOp T\iota_{GO}) \circ T(\widetilde{A}g + \eta_{GO})
            \circ (T\iota_{GAY}\coDiagOp T\iota_{GO}) \circ (\widetilde{A}f + \eta_{GO})
    \\  &\refrel{eq:functor_coDiagOp}=
                \mu_{G(AZ + O)}\circ T(T\iota_{GAZ}\coDiagOp T\iota_{GO})
            \circ \big( T\iota_{TGAZ} \circ T(\widetilde{A}g) \coDiagOp  T\iota_{TGO} \circ T(\eta_{GO}) \big)
            \circ (\widetilde{A}f + \eta_{GO})
    \\  &=
            \mu_{G(AZ + O)}\circ T(T\iota_{GAZ}\coDiagOp T\iota_{GO})
            \circ ( T\iota_{TGAZ}  \coDiagOp  T\iota_{TGO}  )
            \circ ( T(\widetilde{A}g) \times T\eta_{GO} )
            \circ (\widetilde{A}f + \eta_{GO})
    \\  &\refrel{eq:functor_coproduct}=
                \mu_{G(AZ + O)}\circ \big(TT\iota_{GAZ}  \coDiagOp  TT\iota_{GO}\big)
            \circ ( T(\widetilde{A}g) \times T\eta_{GO} )
            \circ (\widetilde{A}f + \eta_{GO})
    \\  &=
                \mu_{G(AZ + O)}
            \circ \big( TT\iota_{GAZ} \circ T(\widetilde{A}(g) \circ \widetilde{A}f
                \coDiagOp  TT\iota_{GO} \circ T\eta_{GO} \circ \eta_{GO} \big)
    \\  &\refrel{eq:coproduct_mult}=
                    T\iota_{GAZ} \circ \mu_{GAZ} \circ T(\widetilde{A}(g)) \circ \widetilde{A}f
                \coDiagOp  T\iota_{GO} \circ \mu_{GO} \circ  T\eta_{GO} \circ \eta_{GO}
    \\  &\refrel[\textbf{A3}, ]{tikz:rightUnit}=
             T\iota_{GAZ} \circ \widetilde{A}(g \circ f)
            \coDiagOp  T\iota_{GO} \circ \eta_{GO}
    \\  &\refrel{eq:MachineEndo_def}=
            \widehat B (g \circ f)
    \text.
  \end{align*}
  Alternatively, the preservation of composition follows from the following commutative diagram:

  \scalebox{0.8}{
  \begin{tikzcd}[ampersand replacement=\&]
  G(AX+O) \ar[r,"\widehat Bf"] \ar[dd,equal]\& TG(AY + O) \ar[d,equal]  \\
  \& T(GAY + GO) \ar[r,"T(\widetilde Ag + \eta_{GO})"] \& [1.5em] T(TGAZ + TGO)) \ar[r,hookrightarrow,"T(\iota_{TGAZ}\coDiagOp\iota_{TGO})"]
  \&[2.1em] TT(GAZ+GO) \ar[r,"\mu_{GAZ+GO}"] \& T(GAZ+GO)\\
  GAX+GO \ar[r,"\widetilde Af + \eta_{GO}"] \&
  \ar[u,hookrightarrow,"\iota_{TGAY}\coDiagOp \iota_{TGO}"]
  TGAY +TGO \ar[r,"T\widetilde Ag + \eta_{TGO}"]\& \ar[u,hookrightarrow,"\iota_{TTGAZ}\coDiagOp\iota_{TTGO}"]TTGAZ + TTGO \ar[ru,hookrightarrow,sloped,"TT(\iota_{GAZ}\coDiagOp\iota_GO)"] \ar[rr,"\mu_{GAZ}+\mu_{GO}"]\&\& TGAZ+TGO \ar[u,hookrightarrow,"\iota_{TGAZ}+\iota_{TGO}"]
  \end{tikzcd}}

  Finally, we check the lifting property by taking some morphism $f\colon X \to Y$ in $\cat C$ and observing
  \begin{align*}
      \widehat{B}\circ L (f)
        &=  \widehat{B} ( \eta_{GY} \circ Gf )
    \\  &\refrel{eq:MachineEndo_def}=
            (TG\iota_{AY} \coDiagOp  TG\iota_O) \circ ( \widetilde A ( \eta_{GY} \circ Gf ) + \eta_{GO} )
    \\  &\stackrel{\mathllap{\textbf{A3}}}=
            (TG\iota_{AY} \coDiagOp  TG\iota_O) \circ ( ( \eta_{GAY} \circ GAf ) + \eta_{GO} )
    \\  &=  (TG\iota_{AY} \circ \eta_{GAY} \coDiagOp  TG\iota_O \circ \eta_{GO} ) \circ ( GAf + \id{GO} )
    \\  &\refrel{eq:coproduct_unit}=
            \eta_{GAY + GO} \circ ( GAf + \id{GO} )
    \\  &=  \eta_{GAY + GO} \circ ( GAf + G(\id{O}) )
    \\  &\stackrel{\mathllap{\textbf{A1}}}=
            \eta_{GAY + GO} \circ ( G(Af + \id{O}) )
    \\  &=%\stackrel{\mathllap{\cref{eq:relKleisli_embedding}}}=
            L B(f)
    \text.
  \end{align*}
\end{proof}

\begin{theoremrep}\label{thm:finalcoal-machine}
Let $T$ be a monad on $\cat C$ and $G$ an endofunctor on $\cat C$.
\begin{enumerate}
	\item If $G$ preserves colimits and the initial algebra $h\colon B(\mu_B) \xrightarrow{\cong} \mu_B$ of $B$ exists in $\cat C$, then $Lh \colon \widehat B (\mu_B) \xrightarrow{\cong} \mu_B$ is the initial algebra of $\widehat B$ in $\klcat (\gmonad T)$.
	\item If $\klcat (T)$ is $\cppo$-enriched, $\overline A$ is locally continuous, and the operation $\_ + g$ commutes with the $\omega$-directed joins, i.e.\ for any increasing families of arrows $(f_i\in\klcat (T)(X,Z))_{i\in\nat}$ we have
	\[
	\bigvee_{i\in\nat} (f_i + g ) = (\bigvee_{i\in\nat} f_i ) + g,
	\]
	then $\widehat B$ is locally continuous.
\end{enumerate}
As a result, $\mu_B$ is the final coalgebra of $\inv{(Lh)} \colon \mu_B \to \widehat B(\mu_B)$ of $\widehat B$ in $\klcat (\gmonad T)$.
\end{theoremrep}
\begin{proof}
For the first statement, it is well-known that $\mu_B$ (if it exists) is the colimit of the sequence in $\cat C$
\[0 \to B0 \to BB0 \cdots\]
And  from \cite[Theorem~2.13]{ULMER1968:DenseAndRelAdjoint} and since $G$ preserves all colimits, we know that $L\mu_B = \mu_B$ is the colimit of the sequence in $\klcat (\gmonad T)$.
\[
0 \to \widehat B0 \to \widehat B\widehat B(0) \cdots
\]

For the second statement, consider the derivation (below $i$ is ranged over $\nat$)
\begin{align*}
\bigvee_{i} \widehat B(f_i) =&\ \bigvee_{i} (TG\iota_{AY} \coDiagOp TG\iota_{C}) \circ (\widetilde A f_i + \eta_{GC})\\
=&\ (TG\iota_{AY} \coDiagOp TG\iota_{C}) \circ \bigvee_{i} (\widetilde A f_i+ \eta_{GC})\\
=&\  (TG\iota_{AY} \coDiagOp TG\iota_{C}) \circ \Big( \big(\bigvee_i \widetilde A f_i \big) + \eta_{GC}\Big) && (\because \text{$\overline A$ is locally continuous})\\
=&\ (TG\iota_{AY} \coDiagOp TG\iota_{C}) \circ \Big(  \widetilde A \big(\bigvee_i f_i \big) + \eta_{GC}\Big)\\
=&\ \widetilde B \big( \bigvee_{i} f_i\big).
\end{align*}
\end{proof}

\section{On constructing a Kleisli lifting}\label{sec:Kl-lift}
In the previous section and in the context of endofunctor $B=\mathbb{A}\times\_ + O$,
we reduced the problem of defining a Kleisli lifting of $A = \mathbb{A}\times\_ $ w.r.t.\thinspace
a relative monad $\gmonad T$ by simply defining the Kleisli lifting of $A$
w.r.t.\thinspace a monad $T$ (cf.\  \ref{assum:barAExists}). The remaining assumptions
\ref{assum:GpresCoproducts} and \ref{assum:GpresA} are straightforward to satisfy
in our case studies (cf.\ \cref{sec:casestudy-cts}). The objective of this section
is to give a general recipe to construct a Kleisli lifting $\overline F \colon \klcat (T)
\to \klcat (T)$ of an endofunctor $F\colon \cat C \to \cat C$. And recall from
\cite{JacobsEtal2015:trace-det} that a Kleisli lifting $\overline F$ is in one-to-one
correspondence with a \kllaw{} $\vartheta\colon FT \Rightarrow TF$. The rest of this section
is devoted to define such a \kllaw{} $\vartheta$ internally using the framework of indexed
categories/fibrations.

To motivate our assumptions that follow, consider the Kleisli category $\klcat (\power)$
induced by the powerset monad. It is well known that the set of Kleisli arrows $\set(X,\power Y)$
are in one-to-one correspondence with the set $\power (X \times Y)$ of binary relations, i.e.
\[
\set (X,\power Y)\ \cong \ \power (X \times Y )\  \cong \ \set (X \times Y, 2),
\]
where $2$ is a two element set \{0,1\}. However, in the Kleisli category $\klcat (\down)$ for the downset
monad $\down\colon \pos \to \pos$ over the category of posets, this idea of representing a Kleisli
arrow as homming into 2 (ordered by the smallest poset generated by the relation $\{(0,1)\}$) is subtly different (see Proposition~\ref{prop:theta-pos}):
\[
\pos(X,\down Y) \cong \{R \subseteq X \times Y \mid R \ \text{is up (down) closed in the first (second) argument}\} \cong \pos (X \times Y^o, 2),
\]
where $Y^o$ denotes the dual poset of $Y$. Thus, in general, we require an endofunctor
$I\colon \cat C \to \cat C$ such that the arrows $X\to TY$ can be internally represented as fibres
 $\Phi(X \times IY)$ in an indexed category.

\begin{enumerate}[label=\textbf{A3.\arabic*},leftmargin=1.1cm]
 %\setcounter{enumi}{1}
% \item\label{assum:writerComonad} Our working category $\cat C$ has products and for a fixed object $\condset \in \cat C$ the following adjoint situation holds $\_ \times \condset
%\dashv \_^\condset$.
 \item\label{assum:icat}  There is an indexed category $\Phi \colon \op{\cat C} \to \pos$  with a bifibration structure,
 %There is an object $\Omega\in\cat C$ (called, henceforth, \emph{truth value} object) that induces an indexed category $\Phi \colon \op{\cat C} \to \pos$ with $\Phi = \cat C(\_,\Omega)$.
%Moreover, $\Phi$ has a bifibration structure,
i.e.\ for each $f\colon X \to Y \in \cat C$ there is an adjoint situation $\exists_f \dashv f^* \colon \Phi X \to \Phi Y$. Note it is customary to write $\Phi f$ as $f^*$ (cf.\thinspace \cite{book:catlogic}).
\item\label{assum:involutive} There is an endofunctor $I\colon\cat C \to \cat C$  such that $F\circ I=I\circ F$.
\item\label{assum:Texistence} There is a monad $(T,\eta,\mu)$ on $\cat C$  with the following correspondence
    \[\theta_{X,Y}\colon \cat C(X,TY) \cong \Phi(X \times IY )\qquad (\text{for each $X,Y\in\cat C$})\] such that the following diagrams commute for each $f\colon X \to X',g\colon Y \to Y'\in\cat C$.
      \[
    \begin{tikzcd}
      \cat C(X,TY) \arrow[r,"\theta_{X,Y}"]
            \arrow[d, phantom, bend right, "\textstyle\textnormal{\textbf{A3.3a}}" left, shift right=3em]
        & \Phi(X\times IY)
      &[6em] \cat{C}(X,TY) \arrow[d,"Tg \circ \_"]\arrow[r,"\theta_{X,Y}"]
      \arrow[d, phantom, bend right, "\textstyle\textnormal{\textbf{A3.3b}}" left, shift right=3em]
      & \Phi(X\times IY) \arrow[d,"\exists_{(X\times Ig)}"]\\
      \cat C(X',TY) \arrow[u,"\_\circ f"] \arrow[r,"\theta_{X',Y}"] & \arrow[u,"(f\times IY)^*"]\Phi(X'\times IY) & \cat{C}(X,TY') \arrow[r,"\theta_{X,Y'}"] & \Phi(X\times IY')
    \end{tikzcd}
    \]
 \item\label{assum:plift} There is an indexed morphism (aka predicate liftings) $\sigma \colon \Phi \Rightarrow \Phi F$. %There is an evaluation map $ev\colon F\Omega \to \Omega$, which are often used in coalgebraic modal logics to define semantics of a modality \cite{Schroeder05:Expressivity,Beohar_et_al:LIPIcs.STACS.2024.10}. Notice that this induces an indexed morphism (aka predicate liftings) $\sigma \colon \Phi \to \Phi F$ defined as $\sigma_X(p)=ev\circ Fp$ (for each $p\in\Phi X$).
\end{enumerate}
Note that our technical objective of this section is to show that Assumptions~\ref{assum:icat}-\ref{assum:plift} imply Assumption~\ref{assum:barAExists}; hence the use of nesting in the above naming convention.

Some remarks are in order on the indexed category $\Phi$. Intuitively, in our case-studies, \ref{assum:icat} will be the fibres $\Phi (X)$ containing predicates of type $X \to \Omega \in \cat C$ with $\Omega\in\cat C$ modelling the truth value object. For our purposes $\Omega=2$ the two-pointed set with the order $0<1$ (when $\cat C=\pos$). The left adjoint $\exists_f$ of $f^*$ is used in categorical logic \cite{book:catlogic} to model the existential quantifier, which is originally due to Lawvere \cite{Lawvere:adjointness}. Below we will use such left adjoints to construct a relation lifting from the predicate lifting $\sigma$, which are  used to define semantics of a modality in coalgebraic modal logics.

The general idea is to define a \kllaw{} $\vartheta \colon FT \Rightarrow TF$ internally using the language of fibred categories. For instance, thanks to the isomorphism in \ref{assum:Texistence}, $\vartheta_X\colon FTX \to TFX$ (for some $X\in\cat C$) can be defined internally by an element in the fibre $\Phi (FTX \times IFX)$. To this end, we start by the identity arrow $\id{TX}$ and consider the element called `membership relation'
\[\inObj X  \ \stackrel{\text{def}} =\ \theta_{TX,X}(\id{TX}) \qquad \in\ \Phi(TX\times IX).\]
This notation is motivated by one of the case studies: when $\cat C=\set$, $I=\text{Id}$, and $T=\power$, this element will correspond to the element relation $\inObj X \subseteq TX\times X$.

Note that technically $\inObj X $ is an element in the fibre $\Phi(TX \times IX)$. Now applying the predicate lifting $\sigma$ on $\inObj X $ gives an element
$
\sigma_{TX \times IX} (\inObj X) \in \Phi F(TX \times IX)
$.
%More generally, define for any relation $R\colon X \times IY \to \Omega$ the operation $\sigma_{X,Y}$ by
%\begin{equation}
%        \label{eq:sigma_def}
%    \sigma_{X,Y}(R) = \proj_\Omega \circ (A(R))
%\end{equation}
We are now a step away from defining our distributive law. First recall the following arrows defined thanks to the universal property of product and the equation $I\circ F=F\circ I$ (cf.\  \ref{assum:involutive}):
\[
 \lambda_{X, Y}=
       F(\proj_X)\diagOp  F(\proj_{IY}) \colon F(X \times IY) \to FX \times FIY = FX \times IFY
\]
where $f \diagOp  f'$, the ``diagonal operation'', is, in general, defined for any two arrows $ f\colon X \to Y $ and $ f\colon X \to Y' $ in $\cat C$ with common domain $X$.
It is the dual to $\coDiagOp$ introduced in connection with \cref{eq:MachineEndo_def}.
%\begin{align*}
%        \lambda_{X, Y}
%    &=
%        A\proj_X\times A\proj_{IY}\colon A(X \times IY) \to AX \times AYX = AX \times IAY
%\intertext{and}
%    \lambda_X
%    &=
%        \lambda_{TX, X}
%\text.
%\end{align*}
Our distributive law $\vartheta_X$ is simply the mapping of $\sigma_{TX\times IX} (\inObj X )$ by the map
$\inv{\theta}_{FTX,FX} \circ
\exists_{\lambda_{TX,X}} \colon \Phi F(TX\times IX) \to \Phi(FTX \times IFX) \to \cat C(FTX,TFX);$
thus, we define by composition of maps
\begin{equation}\label{eq:dlaw-def}
\vartheta_X \stackrel{\mathclap{\text{def}}}
=
	\inv{\theta}_{FTX,FX} \circ \exists_{\lambda_{TX,X}} \circ \sigma_{TX\times IX} \left(\inObj X \right)
=
        \inv{\theta}_{FTX,FX} \circ \exists_{\lambda_{TX,X}} \circ \sigma_{TX\times IX} \circ \theta_{TX,X} ( \id{TX} ).
\end{equation}
%
%\begin{align}\label{eq:dlaw-def}
%  \vartheta_X &\stackrel{\mathclap{\text{def}}}= \ \inv{\theta}_{ATX,AX} \circ \exists_{\lambda_X} \left(\proj_\Omega \circ A(\inObj X )\right)
%\\
%        \label{eq:distrLaw_sigma}
%    &=
%        \inv{\theta}_{ATX,AX} \circ \exists_{\lambda_X} \circ \sigma_{TX, X} \circ \theta ( \id{TX} )
%\end{align}
\begin{lemmarep}
  Let $f\colon X \to Y \in\cat C$. Then we have
  $
  \exists_{TX \times If} (\inObj X ) = (Tf \times IY)^* (\inObj Y ).
  $
\end{lemmarep}
\begin{proof}
Consider the following diagram:
  \[
  \begin{tikzcd}
    \cat C (TX,TX) \arrow[r,"\theta_{TX,X}"] \arrow[d,"Tf\circ\_"']& \Phi(TX \times IX) \arrow[d,"\exists_{TX \times If}"]\\
    \cat C (TX, TY) \arrow[r,"\theta_{TX,Y}"] & \Phi (TX \times IY)\\
    \cat C(TY,TY) \arrow[u,"(Tf)^*"] \arrow[r,"\theta_{TY,Y}"] & \arrow[u,"(Tf\times IY)^*"']\Phi(TY \times IY)
  \end{tikzcd}
  \]
  Note that bottom square commutes since reindexing operation is functorial; whilst, the top one commutes due to \ref{assum:Texistence}. Now the result follows from the fact
 $Tf \circ \id{TX} = Tf = (Tf)^* \id{TY}$.
\end{proof}
As a result, we get the following result which is one of the basic requirements for $\vartheta$ to be a \kllaw{}.
\begin{proposition}\label[proposition]{prop:dlaw-natural}
The map $\vartheta$ defined in \eqref{eq:dlaw-def} is a natural transformation of type $FT \Rightarrow TF$.
\end{proposition}

\subsection*{On relation lifting}
Before we establish the compatibility of $\vartheta$ with unit $\eta$ and multiplication $\mu$ of the monad $T$, respectively, in the subsequent subsections, we need that the mapping
\[
\Phi (X \times IY) \xrightarrow {\sigma_{X\times IY}} \Phi F (X \times IY)
	\xrightarrow{\exists_{\lambda_{X,IY}}}  \Phi (FX \times IFY)
\]
is a relation lifting in the following sense. Notice that, in the sequel, the fibres $\Phi (X \times IY)$ over the object $X\times IY\in\cat C$ are maps of type $X \times IY \to \Omega \in \cat C$, which we simply call as relations. Thus, the above map abbreviated $\tilde\sigma_{X,Y} = \exists_{\lambda_{X,IY}} \circ \sigma_{X\times IY}$ is a candidate to what is known as \emph{$F$-relators} in the literature on coalgebras. Actually, there is no common consensus on the `categorical' definition of an $F$-relator (cf.\  \cite[Chapter~4]{jacobs_coalgBook}), however, for our purpose we require a relator to be an indexed morphism, i.e.\ a natural transformation of type $\Phi(\_ \times IX) \Rightarrow \Phi (\_ \times IFX)$ (for every $X\in\cat C$). Thus, in the parlance of relators, our natural transformation $\vartheta$ on a component is nothing but a relation lifting of the membership relation $\inObj{}$.

\begin{definition}
\begin{subequations}
Given a commuting diagram below on the left, we say the Beck-Chevalley condition for \eqref{eq:square} holds iff the square  \eqref{eq:squareBC} on the right commutes.
\begin{center}
\begin{tikzcd}[ampersand replacement=\&]
            X \arrow[r, "f"] \arrow[d, "g"] \& Y \arrow[d, "k"]    \\
            Z \arrow[r, "h"]                \& Z'
 \end{tikzcd}\quad\eqnum\label{eq:square}
\qquad\qquad\qquad
 \begin{tikzcd}
    \Phi(X) \arrow[r, "\exists_f"]   & \Phi(Y)   \\
        \Phi(Z) \arrow[u, "g^*"] \arrow[r, "\exists_h"]
    &   \Phi(Z') \arrow[u, "k^*"]
\end{tikzcd}\quad\eqnum\label{eq:squareBC}
\end{center}
 \end{subequations}
\end{definition}

\begin{lemmarep}
If in \eqref{eq:squareBC} $k^* \circ \exists_h \leq \exists_f \circ g^*$, then the Beck-Chevalley condition holds for \eqref{eq:square}.
\end{lemmarep}
\begin{proof}
Recall that adjoint $F \dashv G\colon A \to B$ between posets (also called Galois connection) is characterised by $ F(a) \leq b \iff a \leq G(b) $ for $a \in A$, $ b \in B $ and implies $FG(a) \leq a$ and $a \leq GF(a)$.
    Moreover recall that adjoints compose, i.e.\ for further $F' \dashv G'\colon B \to C$ we have $(F' \circ F) \dashv (G \circ G')\colon A \to C $.
    Let the variable $R \in F(Z)$ denote a predicate on $Z$.
    We verify the commutativity:
    {\allowdisplaybreaks
    \[
        \infer=
                {k^* \exists_h (R) = \exists_f g^* (R)}
        {
            \infer=[\exists_f \dashv f^* ]
                {\exists_f g^* (R) \leq k^* \exists_h (R) }
                {
            \infer=[]
                {g^* (R) \leq f^* k^* \exists_h (R) }
                {
            \infer=[\cref{eq:square}]
                {g^* (R) \leq (k \circ f)^* \exists_h (R) }
                {
            \infer=[]
                {g^* (R) \leq (h \circ g)^* \exists_h (R) }
                {
            \infer=[\exists_g \dashv g^*]
                {g^* (R) \leq g^* h^* \exists_h (R) }
                {
            \infer
                { \exists_g g^* (R) \leq h^* \exists_h (R) }
                {   \exists_g g^* (R) \leq R & R \leq h^* \exists_h (R)
                }}}}}}
        &
            \infer=
                {k^* \exists_h (R) \leq \exists_f g^* (R)}
                {\text{assumption}}
        }
    \]
    }
\end{proof}

\begin{corollaryrep}
        \label{cor:pred_preserving_commSquare}
    Let $\cat C$ be $\set$ or $\pos$ and $\Omega=2$ with the order generated by $0<1$ when $\cat C=\pos$. If $\Phi$ is  the indexed category of predicates, i.e.\ $\Phi = \cat C(\_,\Omega)$, then the Beck-Chevalley condition holds for the square in \eqref{eq:square} whenever it is a weak pullback square.
\end{corollaryrep}
\begin{proof}
    %For both categories we have to verify equation \cref{eq:prop:sq_pres_ass:weakPullback}.
    Given a predicate, i.e.\ subset $C$ of $Z$, and take some $y \in k^* \exists_h (C)$.
    Choose some $z \in C$ such that $h(z) = k(y)$.
    Using the weak pullback property, this is to say, there is some $x \in X$ that is mapped to both $y$ via $f$ and $z$ via $g$ (take morphisms from the one element object to find such an $x$).
    Thus $ y \in \exists_f g^*(C) $.
    As $y \in k^* \exists_h (C)$ was chosen arbitrarily, $ k^* \exists_h (C) \subseteq  \exists_f g^*(C)$.
\end{proof}

% \begin{theorem}\label{thm:rlift}
% If the Beck-Chevalley conditions holds for the square in  \cref{eg:BC-lambda} (for any $f$) then the map $\tilde\sigma_{\_,Y}\colon \Phi (\_ \times IY) \Rightarrow \Phi (F\_ \times IFY)$ (for each $Y\in\cat C$) is a natural transformation (i.e.\ a relation lifting). In the context of $\cat C$ is $\set$ or $\pos$ and $\Omega=2$, the Beck-Chevalley condition holds for the above square whenever $F$ preserves weak pullback squares.
% \end{theorem}
\begin{theoremrep}\label{thm:rlift}
\begin{subequations}
The square \cref{eg:BC-lambda_general} below commutes for any $f\colon X \to X'$ and $g\colon Y \to Y'$.
\begin{center}
  \begin{tikzcd}[ampersand replacement=\&]
          F(X \times IY)
              \ar[r, "\lambda_{X,Y}"]
              \ar[d, "F(f \times Ig)"]
      \&   FX \times FIY
              \ar[d, "Ff \times FIg"]
  \\
          F(X' \times IY')
              \ar[r, "\lambda_{X',Y'}"']
      \&   FX' \times FIY'
  \end{tikzcd}\quad\eqnum\label{eg:BC-lambda_general}
\qquad\qquad
\begin{tikzcd}[ampersand replacement=\&]
           F(X \times IY)  \arrow[d,"F(f\times IY)"'] \arrow[r,"\lambda_{X,Y}"] \&
   FX \times IFY \arrow[d,"Ff\times IFY"]\\
   F(X' \times IY) \arrow[r,"\lambda_{X',Y}"'] \&
   FX' \times IFY
\end{tikzcd}\quad\eqnum\label{eg:BC-lambda}
\end{center}
Moreover:
\begin{enumerate}
    \item\label{lst:thm:rlift_BC}
        If the Beck-Chevalley condition holds in the special cases of \cref{eg:BC-lambda} (for any $f$)
        then the map $\tilde\sigma_{\_,Y}\colon \Phi (\_ \times IY) \Rightarrow \Phi (F\_ \times IFY)$ (for each $Y\in\cat C$) is a natural transformation (i.e.\ a relation lifting).
    \item\label{lst:thm:rlift_weakPullback}
        If $F$ preserves weak-pullbacks, then \cref{eg:BC-lambda_general} is a weak pullback-square.
    \item\label{lst:thm:rlift_SetPos}
        In the context of $\cat C$ is $\set$ or $\pos$ and $\Omega=2$, the Beck-Chevalley condition holds for the above square whenever $F$ preserves weak pullback squares.
\end{enumerate}
\end{subequations}
\end{theoremrep}

\begin{proof}
    For the initial claim regarding \cref{eg:BC-lambda_general}, first,
  dualise \cref{eq:functor_coDiagOp} obtaining that for any endofunctor $F$ on $\cat C$ and
  for any $ f\colon X \to X' $ and $ g\colon Y \to Y' $ in $\cat C$
  \begin{equation}\label{eq:functor_diagOp}
          (F {\proj_{X'}} \diagOp  F{{\proj_{Y'}}} ) \circ F( f \times g )
      =   (F(f) \circ {\proj_{X}} ) \diagOp  (F(g) \circ {\proj_{Y}} )
  \text.
  \end{equation}
  Now \cref{eg:BC-lambda_general} commutes, as
  \begin{align*}
  \nonumber
          \lambda_{X',Y'} \circ F(f \times Ig)
        &=   (F({\proj_{X'}}) \diagOp  F({\proj_{IY'}})) \circ F(f \times Ig)
    \\
        &\refrel{eq:functor_diagOp}=
            (F(f) \circ {\proj_{X}} ) \diagOp  (F(Ig) \circ {\proj_{IY}})
    \\  \nonumber
        &=  (Ff \times FIg) \circ (F({\proj_X}) \diagOp  F({\proj_{IY}}))
    \\  \nonumber
        &=  F(f \times Ig) \circ \lambda_{X,Y}
    \text.
  \end{align*}

   For claim~(\ref{lst:thm:rlift_BC}), first, note that by the Beck-Chevalley assumption $\exists_{\lambda_{X, Y}}$ is actually a natural transformation.
   Thus so is $\tilde\sigma_{\_,Y}$ by composition of natural transformations.

   For claim~(\ref{lst:thm:rlift_weakPullback}) we, first, observe that
   \[
   \begin{tikzcd}
            X \times IY \ar[r, "\id{X \times IY}"] \ar[d, "f \times Ig"]
       &    X \times IY \ar[d, "f \times Ig"] \\
       X' \times IY' \ar[r, "\id{X' \times IY'}"]
       &    X' \times IY'
   \end{tikzcd}
   \]
   is a pullback diagram.
   By assumption, so is the left square in
    \[
   \begin{tikzcd}[column sep=huge]
            F(X \times IY) \ar[r, "F(\id{X \times IY})"] \ar[d, "F(f \times Ig)"]
       &    F(X \times IY) \ar[d, "F(f \times Ig)"]
                \ar[r, "F({\proj_X}) \diagOp  F({\proj_{IY}})"']
       &[2em] F(X) \times F(IY)   \ar[l, "i"', bend right=10]
                \ar[d, "Ff \times FIg"]
       &
    \\
       F(X' \times IY') \ar[r, "F(\id{X' \times IY'})"]
       &    F(X' \times IY')
                \ar[r, "F({\proj_{X'}}) \diagOp  F({\proj_{IY'}})"']
       &    F(X') \times F(IY') \ar[l, "i'"', bend right=10]
       &
    \\
        & & & Z \ar[uul, "\pi_R", bend right] \ar[lllu, "\pi_R", bend left=10] \ar[ul, "\pi"]
   \end{tikzcd}
   \]
   which we augmented with a further square as shown.
   Note that the morphisms $i$ and $i'$ are chosen by the weak limit properties of $F(X \times IY)$ and $ F(X' \times IY') $.
   They are not unique but retracts due to the universal product properties of $F(X) \times F(IY)$ and $ F(X') \times F(IY') $.
   The outer square is nothing but \cref{eg:BC-lambda_general}.
   To confirm, that it is a weak pullback, take a test cone with top $Z$ as shown in the diagram.
   From the weak-pullback property of the left square and the maps $ \pi_L $, $ i' \circ \pi $ and $ i \circ \pi_R $ we obtain a map $h\colon Z \to F(X \times IY)$.
   It remains to check that $h$ satisfies the weak universality property for \cref{eg:BC-lambda_general}:
   Obviously, $\pi_L = F(f \times Ig) \circ h$. But also $
        (F({\proj_X}) \diagOp  F({\proj_{IY}})) \circ h
     =  (F({\proj_X}) \diagOp  F({\proj_{IY}})) \circ (i \circ \pi_R)
     =  \pi_R
   $ as $ i $ is a retract.

   Finally, claim~(\ref{lst:thm:rlift_SetPos}) follows from (\ref{lst:thm:rlift_weakPullback}) and \cref{cor:pred_preserving_commSquare}.
\end{proof}

\subsection*{On compatibility of $\vartheta$ with $\eta$ and $\mu$}
It turns out that these compatibility properties, i.e.\ Equations~\eqref{eq:unit-wellbehaved} and \eqref{eq:mult-wellbehaved}, are intrinsically related with Kleisli compositions. Note that because of \ref{assum:Texistence} we can define the composition $\odot$ of relations of type:
\begin{equation}\label{eq:relCompo_def}
    \odot\colon \Phi (Y\times IZ) \times \Phi(X\times IY) \to \Phi (X \times IZ) \qquad S\odot R = \theta_{X,Z}\left(\inv\theta_{Y,Z} (S)\bullet \inv\theta_{X,Y} (R)\right)
\end{equation}
\begin{proposition}
The identity relation $\Delta_X \stackrel{\smash{\text{def}}} = \theta_{X,X}(\eta_X)$ is the unit to the relational composition $\odot$.
\end{proposition}

\begin{lemma}\label{lem:unit-wellbehaved}
If the Beck-Chevalley conditions holds for the squares in \ref{eg:BC-lambda} (for every $f\in\cat C$) then the relation lifting $\tilde\sigma$ preserves the identity relation, i.e.\ $\tilde\sigma_{X,X} (\Delta_X) = \Delta_{FX}$ (for each $X\in\cat C$) if, and only if, $\vartheta$ is compatible with the unit $\eta$ (i.e.\ \cref{eq:unit-wellbehaved} holds).
\end{lemma}
\begin{proof}First observe that
    \[
        \infer=[\text{Def. of $\Delta$}]
        {\infer=
            {\inv\theta_{FX, FX}\tilde\sigma_{X,X} \theta_{X, X} (\eta_X) = \eta_{FX}}
            {\tilde\sigma_{X,X} \theta_{X, X} ( \eta_X ) = \theta_{FX, FX} ( \eta_{FX} )
        }}
        {\tilde\sigma_{X,X} (\Delta_X) = \Delta_{FX}}
    \text.
    \]
    Next observe that the diagram
    \[
        \begin{tikzcd}
                \cat C (TX, TX)     \ar[r, "\theta_{TX,X}"]
                    \ar[d, "\eta_X^*"]
            &   \Phi(TX \times IX)  \ar[r, "\tilde\sigma_{TX,X}"]
                    \ar[d, "(\eta_X\times IX)^*"]
            &   \Phi(FTX \times FIX)\ar[r, "\inv\theta_{FTX,FX}"]
                    \ar[d, "(F\eta_X\times FIX)^*"]
            &[1em]   \cat C (FTX, FTX)
                    \ar[d, "(F\eta_X)^*"]
        \\
                \cat C (X, TX)     \ar[r, "\theta_{X,X}"]
            &   \Phi(X \times IX)  \ar[r, "\tilde\sigma_{X,X}"]
            &   \Phi(FX \times FIX)\ar[r, "\inv\theta_{FX,FX}"]
            &   \cat C (FX, TX)
        \end{tikzcd}
    \]
    commutes by \cref{thm:rlift} and Assumption~\ref{assum:Texistence}a.
    Chasing $\id{TX}$ through the diagram gives
    \[
        \infer=[\cref{eq:dlaw-def}]
            { \vartheta_X \circ F\eta_X
                = \inv\theta_{FX, FX}\tilde\sigma_{X,X} \theta_{X, X} (\eta_X) }
            { \big(\inv\theta_{FTX, FX} \circ \tilde\sigma_{TX,X} \circ \theta_{TX, X} (\id{TX})\big) \circ F\eta_X
                =\inv\theta_{FX, FX}\tilde\sigma_{X,X} \theta_{X, X}
                    ({\id{TX}} \circ \eta_X) }
    \text.
    \]
    Combining the observations of this an the preceding paragraph gives the lemma.
\end{proof}

\begin{lemma}\label{lem:mult-wellbehaved}
    Assume that the Beck-Chevalley conditions holds for the squares in \ref{eg:BC-lambda} (for every $f\in\cat C$).
    If the relation lifting $\tilde\sigma$ preserves the relational composition, i.e.\ for each $R\in\Phi(X\times IY),S\in\Phi (Y\times IZ)$
    \begin{equation}\label{eq:relComp_lifting_compatible}
        \tilde\sigma_{X,Z}(S \odot R) = \tilde\sigma_{FY,FZ} (S) \odot \tilde\sigma_{FX,FY}(R)
    \text,
    \end{equation}
    then $\vartheta$ is compatible with the multiplication $\mu$ (i.e.\ \cref{eq:mult-wellbehaved} holds).
    Conversely, if the compatibility with multiplication occurs, then \cref{eq:relComp_lifting_compatible}
     holds at least in the instance $ R = \inObj X $ and $S =  \inObj{TX}$.
\end{lemma}
\begin{proof} First note that using Kleisli composition condition \cref{eq:mult-wellbehaved} can be expressed by
    \begin{equation}\label{eq:lem:mult-wellbehaved}
        \theta_X \circ F\mu_X = \theta_X \bullet \theta_{TX}
    \end{equation}
    for $X$ ranging over all objects in $\cat C$.
    Second, we consider the same diagram as in the proof of Lemma~\ref{lem:unit-wellbehaved} but with the multiplication $\mu$ in place of the unit $\eta$:
    \[
        \begin{tikzcd}
                \cat C (TX, TX)     \ar[r, "\theta_{TX,X}"]
                    \ar[d, "\mu_X^*"]
            &   \Phi(TX \times IX)  \ar[r, "\tilde\sigma_{TX,X}"]
                    \ar[d, "(\mu_X\times IX)^*"]
            &   \Phi(FTX \times FIX)\ar[r, "\inv\theta_{FTX,FX}"]
                    \ar[d, "(F\mu_X\times FIX)^*"]
            &[1em]   \cat C (FTX, FTX)
                    \ar[d, "(F\mu_X)^*"]
        \\
                \cat C (TTX, TX)     \ar[r, "\theta_{TTX,X}"]
            &   \Phi(TTX \times IX)  \ar[r, "\tilde\sigma_{TTX,X}"]
            &   \Phi(FTTX \times FIX)\ar[r, "\inv\theta_{FTTX,FX}"]
            &   \cat C (FTTX, TX)
        \end{tikzcd}
    \]
    Again the diagram commutes due to \cref{thm:rlift} and Assumption~\ref{assum:Texistence}a.
    Again we chase $\id{TX}$ through the diagram.
    We obtain
    $
        \big(\inv\theta_{FTX, FX} \circ \tilde\sigma_{TX,X} \circ \theta_{TX, X} (\id{TX})\big) \circ F\mu_X
                =\inv\theta_{FTTX, FX}\tilde\sigma_{TTX,X} \theta_{TTX, X} \circ (\mu_X \circ {\id{TTX}})
    $
    which becomes using \cref{eq:dlaw-def}
    {\allowdisplaybreaks
    \begin{align*}
        \vartheta_X \circ F\mu_X
        &= \inv\theta_{FTTX, FX}\tilde\sigma_{TTX,X}
                    \theta_{TTX, X} (\mu_X \circ {\id{TTX}})
    \\  &=  \inv\theta_{FTTX, FX}\tilde\sigma_{TTX,X}
                    \theta_{TTX, X} ({\id{TX}} \bullet {\id{TTX}} )
    \\  &=  \inv\theta_{FTTX, FX}\tilde\sigma_{TTX,X}
                    \theta_{TTX, X} (\inv\theta_{TX,X}(\inObj X) \bullet \inv\theta_{TTX,TX}(\inObj {TX}) )
    \\  &\refrel{eq:relCompo_def}=
            \inv\theta_{FTTX, FX}\tilde\sigma_{TTX,X} ( \inObj X \odot \inObj {TX} )
    \\  &\refrel{eq:relComp_lifting_compatible}=
            \inv\theta_{FTTX, FX}
            \big(\tilde\sigma_{TX,X} ( \inObj X)  \odot \tilde\sigma_{TTX,TX} (\inObj {TX}) \big)
    \\  &\refrel{eq:relCompo_def}=
            \inv\theta_{FTX, FX} (\tilde\sigma_{TX,X}  \inObj X) \bullet  \inv\theta_{FTTX, FTX}(\tilde\sigma_{TTX,TX} \inObj {TX})
    \\  &\refrel{eq:dlaw-def}=
            \theta_X \bullet \theta_{TX}
    \text.
    \end{align*}
    }
    This concludes the forward direction by \cref{eq:lem:mult-wellbehaved}. For the converse direction note that in the single instance where \cref{eq:relComp_lifting_compatible} was
 used when $ R = \inObj X $ and $S =  \inObj{TX}$ is an equivalence since
 %used, it was exactly at the instance $ R = \inObj X $ and $S =  \inObj{TX}$ and the step was an equivalence transformation as
$ \theta_{FTTX, FX} $ is bijective.
    All other steps in the last calculation were equivalences as well.
\end{proof}

Now we can state the main result of this section.

\begin{theorem}\label{thm:kl-law-pos-set}
Let $\cat C$ be either $\set$ or $\pos$ and $\Omega=2$. If $F$ preserves weak pullbacks, then the natural transformation $\vartheta$ defined in \cref{eq:dlaw-def} is a \kllaw{}.
\end{theorem}

We end this section by giving an example of \kllaw{} which is a direct consequence of the above theorem.  Moreover, the general results of neither \cite[Lemma~2.4]{HasuoEtal:GenTraceSemantics} nor \cite[Section~4]{JACOBS2004167} are applicable in Example~\ref{ex:quantitative} since our functor $F$ is not  shapely \cite{HasuoEtal:GenTraceSemantics} and $T$ is a generalisation of powerset monad.

\begin{example}\label{ex:quantitative}
We work with the Lawvere quantale\footnote{The Lawvere quantale is given by the poset $([0,1],\geq)$ with the monoidal operation given by truncated addition $\oplus$, i.e. $r\oplus r' = \min(r + r',1)$.} $\Omega$ and let  $T = \mathcal{P}_\Omega$ be the $\Omega$-valued powerset monad \cite[Remark~1.2.3]{hofmann_seal_tholen_2014} on $\set$ defined as $\mathcal{P}_\Omega = \Omega^X$ on objects and as $Tf(g)(y) = \inf_{f(x)=y} g(x)$ (for $f\colon X\to Y$) on arrows.  Its unit $\eta_X\colon X\to\mathcal{P}_\Omega X$ is given by $\eta_X(x)(x') = 0$ if $x=x'$ and $1$ (the empty meet) otherwise. Multiplication $\mu_X\colon \mathcal{P}_\Omega \mathcal{P}_\Omega X \to \mathcal{P}_\Omega X$ is defined as
$\mu_X(G)(x)=\inf_{g\in \mathcal{P}_\Omega X} G(g)\oplus g(x)$. It is not hard to see that a Kleisli arrow $X \to Y \in \klcat (\power_\Omega)$ corresponds to a $\Omega$-valued matrix of dimension $X\times Y$ (i.e.\ the functor $I$ is set to be the identity functor); the latter are known as $\Omega$-valued relations in \cite{BonchiEtal_2023_upto}. The indexed category $\Phi (X\times Y)$ of $\Omega$-valued relations forms a bifibration; the left adjoint $\exists_f$ (for a function $f\colon X\to Y$) is given by $\exists_f(p)(y)=\inf_{fx=y}p(x)$. Moreover, every weak pullback square in $\set$ satisfies the Beck-Chevalley condition in this quantalic context.

Now consider the functor $F=\mathcal D$ the distribution functor and the predicate lifting $\sigma_X \colon \Omega^X \to \Omega^{\mathcal {D} X}$ given by the expectation $\mathbb E_\mu(p)$:
\[\sigma_X(p)(\mu) = \mathbb E_{\mu}(p) = \sum_{x \in X} p(x) \cdot \mu(x) \qquad \text{(for each $\mu\in\mathcal DX$)}\]
(note that the sum is automatically defined as it ranges over non-negative values and is in $\Omega$ due to the assumptions on $p$ and $\mu$).
Furthermore, the relation $\inObj X$ is given by evaluation $(p,x) \mapsto p(x)$ and the natural transformation $\lambda\colon \mathcal D(X\times Y) \to \mathcal D(X) \times \mathcal D(Y)$ maps a joint distribution $\omega \in\mathcal D(X\times Y) \mapsto (\sum_{y\in Y} \omega(\_,y),\sum_{x\in X} \omega(x,\_))$ to its corresponding marginal distributions. Thus, using the terminology of optimal transport, the left adjoint $\exists_{\lambda_X}(M)(\mu,\nu)=\inf_{\lambda_X(\omega)=(\mu,\nu)} M(\omega)$ computes the best possible coupling of a given pair of distributions $\mu,\nu$ in $M$. This gives rise to a \kllaw{} $\vartheta \colon \mathcal D \power_\Omega \to \power_\Omega \mathcal D$ as follows:
\[
\vartheta_X(M)(\mu) = \inf_{\lambda_{TX,X}(\omega) = (M,\mu)} \mathbb E_\omega(\inObj X) = \inf_{\lambda_{\mathcal D X,X}(\omega)
= (M,\mu)} \sum_{(p,x) \in \power_\Omega X \times X } p(x)\cdot \omega(p,x),
\]
where $M\in \mathcal D(\power_\Omega X)$ and $\mu\in \mathcal D(X)$.
\end{example}
%\begin{enumerate}
%\item $c\colon X\to TFX \iff X \to \mathcal D(\power X)$.
%\item $\sigma_X \colon \Omega^X \to \Omega^{\power ( X)}$ given by $\sigma_X(p)(U)=1 \iff \mu(p)>0$.
%\item $\vartheta_X\colon \mathcal D(\power X) \to \power( \mathcal DX)$ given by $\mu\in\vartheta(M) \iff \exists_{x,U}\ \mu(x)>0 \land x\in U \land M(U)>0$.
%\item $\sigma_X(\inObj X)(c)=\sum_{c(p,x)>0} p(x)\cdot c(p,x)$
%\item $\lambda_X\colon \mathcal D(X\times Y) \to \mathcal D(X) \times \mathcal D(Y)$ .
%\item $\exists_{\lambda_{TX,X}}(\sigma(\inObj X))(M,\mu) = \inf_{\lambda(c)=(M,\mu)} \sum_{c(p,x)>0} p(x)\cdot c(p,x)$
%\end{enumerate}
\section{A case study on conditional transition system (CTS)}\label{sec:casestudy-cts}
In this section, we will apply \cref{thm:finalcoal-machine} to synthesise language, failure and ready equivalences for CTSs. CTSs are a generalisation of labelled transitions systems (LTSs) aimed at modelling a family of LTSs in a compact manner; thus, they are suited to formally model a software product line \cite{CTS_withupgrades-TASE17}.

\begin{definition}
A \emph{conditional transition system} (CTS) over an alphabet $\mathbb{A}$ and a finite poset $\condset$ of conditions is a quadruple $(X,\mathbb{A},\condset,\rightarrow)$, where $X$ is a set of states and ${\rightarrow} \subseteq X \times \mathbb{A} \times \condset \times X$ is the transition relation satisfying the following condition (below  we write $x \xrightarrow {a,k} y$ to denote the predicate $(x,a,k,y) \in {\rightarrow}$):
\[
\forall_{x,y\in X,a\in \mathbb{A},k,k'\in \condset}\ (x \xrightarrow{a,k} y \land k'\leq k) \implies x \xrightarrow{a,k'} y.
\]
\end{definition}

The operational intuition behind a CTS with upgrades is as follows. A CTS starts executing its behaviour from a state $x$ and by arbitrarily choosing a condition $k\in\condset$. Note that all the transitions that are enabled at $x$ and are guarded by a condition greater than or equal to $k$ are activated, while the remaining transitions remain inactive. Henceforth, the system behaves like a traditional LTS, though at any point in its evolution the system may upgrade to a condition $k'\leq k$. If the set $\condset$ is trivially ordered, then we call the system as a CTS without upgrades (originally introduced in \cite{CTS-original}).

%
%\todo[inline]{Sebastian: Please add subsections on coalgebraic modelling of CTSs with/without upgrades and apply \cref{thm:finalcoal-machine}}
In the sequel we will fix a set of actions $\mathbb{A}$ that, whenever order is taken into account, is trivially ordered (i.e.\ by equality). For the systems we model, we additionally assume that the state space, usually denoted by $X$, is always trivially ordered. We define the behavioural notions for CTS as follows:
\begin{definition}[Behavioural Equivalences]
Let $(X,\mathbb{A},\condset,\rightarrow)$ be a CTS over $\condset$ then we define:
\begin{itemize}
	\item Assume $\downarrow \subseteq X$ modelling the set of accepting/terminating states, then the $k$-language of a state $x\in X$: $$L(x,k)=\{w\in \fseq{\mathbb{A}}\mid\exists_{x'}\ x' \in X \land x\steps{w,k} x' \land x' \in {\downarrow}\},$$
	where ${\steps{k}}\subseteq X \times \fseq{\mathbb{A}} \times X$ is the usual reachability relation on the state space.
	\item Taking into consideration that upgrades may allow additional steps, we call $x$ and $y$ equivalent for a condition $k\in\condset$ iff $\forall k'\leq k: L (k',x)=L (k',y)$.
	Two states are \emph{conditionally language equivalent}, if they are language equivalent for all $k\in\condset$.
	\item We define the failure pairs, resp.\ ready pairs, of a state $x\in X$ for a condition $k\in\condset$ as:
	\begin{align*}
	F(x,k)=&\ \{(w,U)\mid \exists_{x'\in X}\ x\steps{w,k}x' \ \land\ \forall_{a\in U}\nexists_{x''}\ x'\xrightarrow {a,k} x'' \}\\
	R(x,k)=&\ \{(w,U)\mid \exists_{x'\in X}\ x\steps {w,k}x'\ \land\ \forall_{a\in U}\exists_{x''\in X}\ x'\xrightarrow {a,k}x''\}.
	\end{align*}
	%$$F(x,k)=\{(w,U)\mid \exists x'\in X: x\xrightarrow{w,k}^*x'\wedge \forall a\in U:x'\not\xrightarrow {a,k}\}$$
	\item Taking into consideration that upgrades may allow additional steps we call two states $x,y\in X$ failure (resp.\ ready) equivalent for a condition $k\in\condset$ iff $F(x,k')=F(y,k')$ (resp.\ $R(x,k')=R(y,k')$) for all $k'\leq k$. We call them \emph{conditionally failure equivalent} (resp.\ \emph{conditionally ready equivalent}), if $F(x,k)=F(y,k)$ for all conditions $k\in\condset$.
%	\item Similarly, we define the ready pairs of a state $x\in X$ for a condition $k\in\condset$ as:
%	$$R(x,k)=\{(w,U)\mid \exists x'\in X: x\xrightarrow {w,k}^*x'\wedge\forall a\in U\exists x''\in X:x'\xrightarrow {a,k}x''\}$$
%	\item Taking into consideration that upgrades may allow additional steps we call two states $x,y\in X$ ready equivalent for a condition $k\in\condset$ iff $R(x,k')=R(y,k')$ for all $k'\leq k$. We call them conditionally ready equivalent, if $R(x,k)=R(y,k)$ for all conditions $k\in\condset$.	
\end{itemize}
\end{definition}

The three notions of behaviour we want to model coalgebraically can all be modelled by variations of the functor $\widehat B$ from \cref{thm:finalcoal-machine} either for the Kleisli category induced by the powerset monad $\power $ on \set\ or the downset monad $\power _\downarrow$ on \pos. We can choose the behavioural notion by our choice of the set of observations $O$ and we can vary between systems with and without upgrades by giving $\condset$ an order or no order. The following table shows the modelling choices succinctly:
\begin{center}
\begin{tabular}{ c| c c }
  & $\condset$ unordered & $\condset$ ordered \\ \hline
 $O=\{1\}$ & $\hat o=(k,\bullet)$ if $x\in\downarrow$ & $\hat o=(k,\bullet)$ if $x\in\downarrow$\\
 $O=\power (\mathbb{A})$ (refusal sets) & $\hat o=\{(k,a) \mid  \nexists_{x'}\ x \xrightarrow{k,a}x'\}$ & undefined\\
 $O=\power (\mathbb{A})$ (ready sets) & $\hat o=\{(k,a)\mid \exists_{x'}\ x\xrightarrow{k,a}x'\}$ & $\hat o=\{(k',a)\mid k'\leq k\wedge \exists_{x'}\ x\xrightarrow{k',a}x'\}$
\end{tabular}
\end{center}
%\begin{center}
%\begin{tabular}{ c| c c }
%  & $\condset$ unordered & $\condset$ ordered \\ \hline
% $O=\{1\}$ & Language eq.\ without upgrades & Language eq.\ with upgrades \\
% $O=\power (\mathbb{A})$ (refusal sets) & Failure eq.\ without upgrades & Failure eq.\ with upgrades\footnotemark\\
% $O=\power (\mathbb{A})$ (ready sets) & Ready eq.\ without upgrades & Ready eq.\ with upgrades
%\end{tabular}
%\footnotetext{this variant can only be compute in \set\ because the coalgebras are order-inverting rather than order-preserving.}
%\end{center}
In particular, for a coalgebraic modelling of a CTS $(X,\mathbb A,\condset, \rightarrow)$ (with or without upgrades), we consider the following function
%over an ordered set of conditions $\mathbb K$ and action set $A$, then for ready equivalence, we model the coalgebra as 
$\alpha\colon \condset \times X\to T(\condset\times \mathbb{A}\times X + O)$ where 
\[
\alpha(k,x)=\{(k', a, x')\mid x\xrightarrow{k',a}x' \land k'\leq k\}\cup \{\hat o \in \condset\times O \mid \text{$\hat o$ as defined in the above table}\},
%\alpha(k,x)=\{(k', a, x')\mid x\xrightarrow{k',a}x', k'\leq k\}\cup\{(k',a)\mid k'\leq k\wedge \exists x'\in X\mid x\xrightarrow{k',a}x'\}.
\]
$T=\power,\cat C=\set$ for CTSs without upgrades and $T=\power_\downarrow,\cat C=\pos$ for CTSs with upgrades. In other words, the coalgebra map models local conditional behaviour, i.e.\thinspace it models immediate transitions and immediate ready sets from a state at a particular condition. Note that $\alpha(k,x)$ need \emph{not} be a downward closed set if we incorporate refusal sets in the ordered case like the ready sets,  i.e.\ by including the clause $o=\{(k',a) \mid k'\leq k \land \nexists_{x'}\ x\xrightarrow{k',a}x'\}$ in the definition of $\alpha$. This is because refusal sets are order reversing, unlike ready sets which are order preserving. As a result, we are \emph{unable} to capture conditional failure equivalence coinductively for CTSs with upgrades; but we can without upgrades (cf.\thinspace Theorem~\ref{thm:cts-noupgrades-coinductive}) since $\alpha(k,x)$ (for arbitrary $k\in\condset,x\in X$) is downward closed when $\condset$ is trivially ordered.

%Note that in principle, we could work out all notions in \set, but due to the involved order on $\condset$, it is more natural to consider the system types with upgrades in the category of \pos. So, the system of interests are coalgebras $X \to \mathbb{A}\times X + O = AX + O\in \klcat (\gmonad T)$ where $G=\condset\times\_$ and $\cat C = \set, T=\power$ (or, respectively, $\cat C = \pos, T=\power_\downarrow$). \hbnote{Seb: could you add the definition of our coalgebra map and mention what is the order on $O$ when $\cat C=\pos$?} Thus, 
So the functor $G$ throughout this section is the writer comonad $\condset\times\_$. Moreover, $\condset\times\_ \dashv {\_}^{\condset}$ both in $\set$ and $\pos$, we immediately have that Assumption~\ref{assum:GpresCoproducts} is valid since left adjoints preserve colimits. In addition, Assumption~\ref{assum:GpresA} holds because $GAX = \condset \times \mathbb{A} \times X \cong \mathbb{A} \times \condset\times X = AGX$. The next two subsections are on Assumption~\ref{assum:barAExists}, i.e.\ a Kleisli lifting $\overline A$ exists for the functor $A\colon \cat C \to\cat C$ (given by $AX=\mathbb{A}\times X$ for $X\in\cat C$ and $\cat C \in \{\set,\pos\}$).

\subsection*{CTS without upgrades when $C=\set$ and $T=\power$}
Recall that a Kleisli category $X \to \power Y \in \set$ is isomorphic to a binary relation $X \times Y \to 2$, where $2=\{0,1\}$. Thus, we let $I=\text{Id}$ and $\Phi \colon \op\set \to \pos$ be the indexed category of (Boolean) predicate, i.e. $\Phi X =\set(X,2) \cong \power X$ and $f^*$ is given by the inverse image $\inv f$ (for each $f\in\set$). Moreover, $\Phi$ has the bifibration structure since the left adjoint $\exists_f$ (for a function $f\colon X\to Y \in\set $) is given by the direct image $\exists_f(U) = \{fx \mid x\in U\}$ (for each $U\in\Phi X$). Thus, Assumptions~\ref{assum:icat} and \ref{assum:involutive} hold.
\begin{propositionrep}
With the above definitions of $I$ and $\Phi$, Assumption~\ref{assum:Texistence} is valid.
\end{propositionrep}
\begin{proof}
We have to give an isomorphism \[\theta_{X,Y}\colon \cat C(X,TY) \cong \Phi(X \times IY )\qquad (\text{for each $X,Y\in\cat C$})\] for $T=\power $ and $I=\text{Id}$. For any $f\in C(X,TY)$ we define:
$$\theta_{X,Y}(f)(x,y)=\begin{cases}1&\text{if\ }y\in f(x)\\0&\text{otherwise}\end{cases}.$$
When interpreted as relations, this can be easier written as $$\theta_{X,Y}(f)=\{(x,y)\mid y\in f(x)\}.$$
Note that this definition also works in \pos, because this arrow is trivially order preserving. We now show the necessary identities.
\begin{itemize}
	\item Let $f\colon X\rightarrow X'$ be any arrow in \set. To show $(f\times IY)^*\circ\theta_{X',Y}=\theta_{X,Y}\circ f^*$, let any $h\in\cat C(X',TY)$ be given and compute:
	$$(f\times IY)^*\circ\theta_{X',Y}(h) = (f^{-1}\times Y)(\{(x',y)\mid h(x')\ni y\})=\{(x,y)\mid h(f(x))\ni y\}$$
	and:
	$$\theta_{X,Y}\circ f^*(h)=\{(x,y)\mid\exists x'\in X: x\in f^{-1}(x')\wedge h(x')\ni y\}=\{(x,y)\mid h(f(x))\ni y\}$$
	\item Let $g\colon Y\rightarrow Y'$ be any arrow in \set. To show $\exists_{X\times Ig}\circ\theta_{X,Y}=\theta_{X,Y'}\circ (Tg\circ\_)$, let any $h\in\cat C(X,TY)$ be given and compute:
	$$\exists_{(X\times Ig)}\circ\theta_{X,Y}(h)=\exists_{(X\times Ig)}(\{(x,y)\mid y\in h(x)\})=\{(x,g(y)\mid y\in h(x)\}$$
	and:
	$$\theta_{X,Y'}\circ (Tg\circ\_)(h)=\{(x,y')\mid y'\in\power g\circ h(x)\}=\{(x,y')\mid\exists_{ y\in h(x)} g(y)=y')\}=\{(x,g(y))\mid y\in h(x)\}$$
\end{itemize}
\end{proof}
To satisfy Assumption~\ref{assum:plift}, consider the predicate lifting $\sigma_X \colon \Phi X \to \Phi (AX)$ given by the mapping
\[
U\subseteq X \ \mapsto \ \sigma_X(U) = \{(a,x) \mid x\in U \land a \in \mathbb{A}\} = \mathbb{A} \times U.
\]
Clearly, $\sigma$ is an indexed morphism because $f^*$ is a functor. Now we can compute the relation lifting $\tilde \sigma_{X,Y} \colon \Phi (X \times Y) \to \Phi (AX \times AY)$ to be:
\[
\tilde\sigma_{X,Y} R = \exists_{\lambda_{X,Y}} \sigma_{X\times Y} (R)
% = \exists_{\lambda_{X,Y}} \{(a,x,y) \mid x\mathrel R y \land a\in \mathbb{A}\}
 =  \{\lambda_{X,Y}(a,x,y) \mid x \mathrel R y \land a\in \mathbb{A}\}
 =  \big\{\big( (a,x), (a,y)\big) \mid x\mathrel R y \land a \in \mathbb{A} \big\}.
\]
%{\allowdisplaybreaks
%\begin{align*}
%\tilde\sigma_{X,Y} R =&\ \exists_{\lambda_{X,Y}} \sigma_{X\times Y} (R)\\
% =&\ \exists_{\lambda_{X,Y}} \{(a,x,y) \mid x\mathrel R y \land a\in \mathbb{A}\}\\
% = &\ \{\lambda_{X,Y}(a,x,y) \mid x \mathrel R y \land a\in \mathbb{A}\}\\
% = &\ \Big\{\Big( (a,x), (a,y)\Big) \mid x\mathrel R y \land a \in \mathbb{A} \Big\}.
%\end{align*}
%}
\begin{propositionrep}
The above relation lifting $\tilde\sigma$ preserves identity relations and relational composition.
\end{propositionrep}
\begin{proof}
Consider the identity relation on a subset $U\subseteq X$: $$I=\{(x,x)\mid x\in U\}$$
then we can compute:
$$\sigma_X(I)=\{((a,x),(a,x))\mid x\in U \land a\in \mathbb{A}\}$$
This is the identity relation on $\mathbb{A}\times U$.

Now let $U_1, U_2, U_3\subseteq X$ and relations $R_1\subseteq U_1\times U_2$, $R_2\subseteq U_2\times U_3$ be given, then
\begin{equation*}\begin{aligned}
&\sigma_X(R_2)\circ\sigma_X(R_1)=\{((a,u_2),(a,u_3))\mid(u_2,u_3)\in R_2\}\circ\{((a,u_1),(a,u_2))\mid(u_1,u_2)\in R_1\}\\
=&\{((a,u_1),(a,u_3))\mid\exists u_2\in U_2:(u_1,u_2)\in R_1\wedge(u_2,u_3)\in R_2\}
\end{aligned}\end{equation*}
and
\begin{equation*}\begin{aligned}
&\sigma_X(R_2\circ R_1)=\sigma_X(\{(u_3,u_1)\mid\exists u_2\in U_2: (u_1,u_2)\in R_1\wedge(u_2,u_3)\in R_2\})\\
=&\{((a,u_1),(a,u_3))\mid\exists u_2\in U_2:(u_1,u_2)\in R_1 \wedge(u_2,u_3)\in R_2\}
\end{aligned}\end{equation*}
\end{proof}
And thanks to \cref{thm:kl-law-pos-set}, \cref{eq:dlaw-def} gives a \kllaw{} $\vartheta\colon A\power \Rightarrow \power A $. More concretely, on a component $X\in\set$, it is given as $\vartheta_X=\inv{\theta}_{A\power X,AX} \circ \tilde \sigma_{\power X\times X} (\inObj X) = \inv{\theta}_{A\power X,AX} \{((a,U),(a,x)) \mid x\in U\} $. Thus,
\[\vartheta_X(a,U) =\{(a,x) \mid x\in U\}. \]
So we obtain Kleisli liftings $\overline A\colon\klcat (\power) \to \klcat (\power)$ of $A$ (below $X\in\set$ and $f\colon X\to \power Y \in\set$)
\[
\overline A X= AX=\mathbb{A}\times X \ \text{and}\ \overline Af(a,x) = \{(a,y) \mid y\in fx\}  .
\]
\begin{example}\label{ex:machine-lift-abstractly}
In the above paragraph, let $F\colon \set \to \set$ be $B=\mathbb{A} \times\_ + O$ (instead of just $\mathbb{A}\times \_$). Now the following predicate lifting $\sigma\colon\Phi X \to \Phi FX$ induces a relation lifting $\tilde\sigma\colon\Phi (X\times Y) \to \Phi (FX \times FY)$:
{\allowdisplaybreaks
\begin{align*}
U \subseteq X \mapsto&\ \sigma_X U = O  \cup \{(a,x) \mid x \in U \land a\in \mathbb{A}\}\\
R \subseteq X \times Y \mapsto&\ \tilde \sigma_X U = \Delta_O  \cup \{((a,x),(a,y)) \mid x\mathrel R y \land a\in \mathbb{A}\}\
\end{align*}
}
So from \cref{thm:kl-law-pos-set} we get a \kllaw{} $\vartheta\colon F\power \Rightarrow \power F$, which on a component $X\in\set$ is $\vartheta_X(a,U)=\{a\} \times U$ and $\vartheta_X(o)=\{o\}$. This induces a Kleisli lifting $\overline B$ as $\overline BX= BX$ and $\overline Bf =\vartheta_Y \circ Bf$ (for every $f\in X \to Y \in\klcat (\power)$). The latter coincides with the definition of $\overline Bf$ given in \cref{eq:machine-lifting}.
\end{example}

\cref{thm:KlExtMachineEndo} is now applicable and giving us a Kleisli lifting
$\widehat B \colon \klcat (\gmonad \power) \to \klcat (\gmonad \power)$, which on objects is $\widehat B(X) = BX = \mathbb{A}\times X + O$ (for each $X\in\set$) and on an arrow $f\colon\klcat (\gmonad \power)(X,Y)$ is defined as follows: $\widehat B(f)(k,o)=\{(k,o)\}$ and using \cref{eq:MachineEndo_def} we get
\[
\widehat B(f)(k,a,x) = TG\iota_{AY} \circ \widetilde Af (k,a,x) = \{(k',(a',y)) \mid (k',y)\in f(k,x)\}
\]

The next proposition is a consequence of \cref{thm:finalcoal-machine}.
\begin{propositionrep}\label{prop:Set-CTS-coincide}
The Kleisli lifting $\overline A$ is locally continuous and the operation $\_ + f$ (for every fixed arrow $f\in\klcat (\power)$) on Kleisli homsets commutes with $\omega$-directed joins. As a result, the initial algebra $L(\mu_B) = \fseq{\mathbb{A}} \times O$ coincides with the final coalgebra of $\widehat B$ in $\klcat (\gmonad\power)$.
\end{propositionrep}
\begin{proof}
From \cite{HasuoEtal:GenTraceSemantics} we know that $\klcat (\power)$ is $\cppo$-enriched and $\omega$-directed join is given by union. Now let $f_i$ be an increasing $\omega$-chain of arrows in $\klcat (\power)$, then
\[
\bigcup_{i} \overline Af_i (a,x) = \bigcup_{i} \{(a,y) \mid y\in f_ix\} = \{(a,y) \mid y\in \bigcup_i f_i x\} = \overline A(\bigcup_i f_i)(a,x).
\]
Now let $f_i\in\klcat (\power)(X,Z)$ be a family of Kleisli arrows and let $g\in\klcat (\power) (Y,Z)$. Without loss of generality assume $X,Y$ are disjoint. Then,
\[
(\bigcup_i f_i + g)x = \bigcup_i f_i x = (\bigcup_i f_i + g)x  \quad (\bigcup_i f_i + g)y = gy = (\bigcup_i f_i + g)y.
\]
Now the result follows directly from \cref{thm:finalcoal-machine}.
\end{proof}

\begin{theoremrep}\label{thm:cts-noupgrades-coinductive}
Let $\alpha\colon X \to AX + O \in\klcat(\gmonad \power)$ be a coalgebra.
\begin{enumerate}
\item If $O=1$ then the unique coalgebra homomorphism is given by the mapping $(k,x)\mapsto \{k\} \times L(x,k)$.
\item If $O=\power {\mathbb{A}}$ and $\alpha$ models the ready set (resp.\thinspace refusal set) of a state, i.e.\ $(k,o)\in \alpha(k,x)$ iff $o$ is the set of actions enabled (resp.\thinspace disabled) from the state $x$ at condition $k$, then the unique coalgebra homomorphism is given by the mapping $(k,x) \mapsto \{k\} \times R(x,k)$ (resp.\thinspace $(k,x) \mapsto \{k\} \times F(x,k)$).
\end{enumerate}
\end{theoremrep}

\begin{proof}
First, note that the initial algebra $\mu_B$ of $B$ in $\set$ exists and is given by $h+h'\colon B(\fseq{\mathbb{A}} \times O) \to \fseq{\mathbb{A}}\times O$ as defined in \cref{prop:IAlgExist-Set}. Thus, by \cref{thm:finalcoal-machine} the map $\inv{(L(h+h'))} \colon \mu_B \to \widehat B (\mu_B)$ is the final coalgebra. Note $L(h+h')\colon G(A(\fseq{\mathbb{A}} \times O) + O)\to \power G(\fseq{\mathbb{A}}\times O)$ maps $(k,o) \mapsto \{(k,\epsilon,o)\}$ and $(k,a,w,o) \mapsto \{(k,aw,o)\}$. And since $\mu_B$ and $\widehat B \mu_B$ are isomorphic in $\klcat(\gmonad\power)$ because
\[\mu_B \cong B (\mu_B) \implies \mu_B = L\mu_B \cong LB (\mu_B) = \widehat B L(\mu_B) = \widehat B(\mu_B),\]
so the final coalgebra for $\widehat B$ is $(\fseq{\mathbb{A}}\times C)$ with the mapping $\beta=\inv{(L(h+h'))}$, i.e.
$$\beta\colon(\fseq{\mathbb{A}}\times O)\rightarrow\widehat B(\fseq{\mathbb{A}}\times O)\qquad (k,(aw, o))\mapsto \{(k,a,(w,o))\}\qquad (k,(\epsilon, o))\mapsto \{(k,o)\}$$
we will show, that for $O=1$, language equivalence is the behavioural equivalence and for $O=\power (\mathbb{A})$, ready equivalence is the behavioural equivalence.

We first observe that with $O=1$, the final coalgebra collapses to the carrier $\fseq{\mathbb{A}}$, because $\fseq{\mathbb{A}}\times\{\bullet\}$ is isomorphic to $\fseq{\mathbb{A}}$.

The arrow into the final coalgebra can be defined directly via the language $L$ of a given state, namely
$$f(k,x)=\{(k,w)\mid w\in L(x,k)\}.$$
It is clear that $f(k,x)=f(k,y)$ iff $L(x,k)=L(y,k)$, so if the claim that this is the (unique) arrow into the final coalgebra, is proven, we have shown that language equivalence is captured by
this model. We compute:
\begin{equation*}\begin{aligned}
\mu\circ\power \widehat Bf\circ\alpha(k,x) = & \mu\power \widehat Bf(\{(k,\bullet),(k,a,x')\mid x\xrightarrow{a,k}x'\})\\=&\{(k,\bullet),(k,a,w)\mid \exists_{x'\in X}\ w\in L(x',k)\wedge
x\xrightarrow{a,k}x'\}
\end{aligned}\end{equation*}
and:
\begin{equation*}\begin{aligned}
\mu\circ\power \beta \circ f(k,x) = & \mu\power \beta (\{(k,w)\mid w\in L(x,k)\})\\
=&\{(k,\bullet)\mid \epsilon\in L(x,k)\}\cup\{(k,a,w)\mid aw\in L(x,k)\}\\
=&\{(k,\bullet),(k,a,w)\mid \exists_{x'\in X}\ w\in L(x',k)\wedge x\xrightarrow{a,k}x'\}
\end{aligned}\end{equation*}
So the coalgebra homomorphism diagram commutes, concluding our proof for language equivalence.

For ready equivalence, we have to use $O=\power (\mathbb{A})$ and we consider the following map: $$f\colon X\rightarrow \widehat B(\fseq{\mathbb{A}} \times\power (\mathbb{A}))\qquad f(k,x)=\{k\}\times R(x,k)$$
where $R(x,k)$ for any state $x\in X$  and any condition $k\in K$ is the set of ready pairs starting in $x$ under the condition $k$.

Clearly, two states map to the same point in the final coalgebra iff they are ready equivalent,
so we we only need to show that the arrow $f$ is actually a coalgebra homomorphism (therefore necessarily unique):
\begin{equation*}\begin{aligned}
\mu\circ\power \widehat Bf\circ\alpha(k,x) = & \mu\power \widehat Bf(\{(k,a,x')\mid x\xrightarrow{a,k}x'\}\cup\{(k,U)\mid\forall a\in U\exists x'\in X:
x\xrightarrow{a,k}x'\})\\=&\bigcup\{\{k\}\times\{a\}\times
R(x',k)\mid x\xrightarrow{a,k}x'\}\cup\{(k,U)\mid \forall a\in U\exists x'\in X: x\xrightarrow{a,k}x'\}
\end{aligned}\end{equation*}
and:
\begin{equation*}\begin{aligned}
\mu\circ\power \beta \circ f(k,x) = & \mu\power \beta (\{k\}\times R(x,k))\\
=&\{(k,U)\mid (\epsilon,U)\in R(x,k)\}\cup\{(k,a,(w,U))\mid (aw,U)\in R(x,k)\}\\
=&\{(k,U)\mid \forall a\in U\exists x'\in X: x\xrightarrow{a,k}x'\}\cup\{(k,a,(w,U))\mid \exists x'\in X: x\xrightarrow{a,k}x'\wedge(w,U)\in R(x',k)\}\\
=&\{(k,U)\mid \forall a\in U\exists x'\in X: x\xrightarrow{a,k}x'\}\cup\bigcup\{\{k\}\times\{a\}\times R(x',k)\mid x\xrightarrow{a,k}x'\}
\end{aligned}\end{equation*}
Note that we can characterise failure equivalence completely analogously: We only have to change the map $f$ to:
$$f\colon X\rightarrow \widehat B(\fseq{\mathbb{A}}\times\power (\mathbb{A}))\qquad f(k,x)=\{k\}\times F(x,k)$$where $F(x,k)$
for any state $x\in X$ and any condition $k\in K$ is the set of failure pairs starting in $x$ under the condition $k$. The computation is completely analogous.
\end{proof}

\subsection*{CTSs with upgrades when $C=\pos$ and $T=\power_\downarrow$}
We begin by recalling the downset monad $\power_\downarrow$ on the category $\pos$ of posets that maps a poset to its downward closed subsets. In particular,
\[
\down (X) = \{U \subseteq X \mid U = \history U\} \qquad \history U = \{x' \mid \exists_x\ x\in U \land x'\leq x\}.
\]
On an arrow $f\colon X\to Y \in \pos$, $\power_\downarrow$ maps a downward closed subset $U\subseteq X$ to the downward closed subset $\history f(U)$. Moreover, the unit $\eta_X$ maps a point $x$ to its history $\history \{x\}$ and the multiplication $\mu$ is given by union. Now a Kleisli arrow $f\colon X \to\down Y \in \pos$ and the relation $\theta_{X,Y}(f) = \{(x,y) \mid y\in fx\}$. It is not hard to see that the relation is upward closed in $X$ and downward closed in $Y$ (or alternatively upward closed in the dual poset $Y^o$). Moreover, it is well known that upward closed sets of a poset $X$ are in correspondence with poset maps of type $X \to 2$ where $2=\{0,1\}$ and is ordered by $0<1$.

Due to these considerations, define an indexed category $\Phi \colon \op\pos \to \pos$ of upward closed subsets of a poset and $I\colon \pos \to \pos$ to be the functor that maps a poset to its dual. Moreover, the reindexing $f^*$ (for an arrow $f\in\pos$) is given by the inverse image $\inv f$ (since inverse image preserves upward closed subsets). Moreover, $\Phi$ has the bifibration structure since the left adjoint $\exists_f$ (for $f\colon X\to Y \in \pos$) is given by the set $\exists_f (U) = \{y \in Y \mid \exists_x\ x\in U \land fx \leq y\}$, for each $U\in \Phi X$.
\begin{propositionrep}
With the above definition of $\Phi$, Assumption~\ref{assum:icat} is valid. Moreover, Assumption~\ref{assum:involutive} is also valid when $F=A$.
\end{propositionrep}
\begin{proof}
For a poset arrow $f\colon X \to Y$ we have $\exists_f U \subseteq V \iff U \subseteq f^*V$.

Let $U\subseteq f^* V$ and let $y\in\exists_{f} U$. I.e., $fx \leq y$ for some $x\in U$. Then $x\in f^* V$, i.e.\ $fx \in V$ and since $V$ is upward closed subset of $Y$ we have $y\in V$.

For the converse, let $\exists_f U \subseteq V$ and $x\in U$. Then $fx \in \exists_f U$ and thus $fx\in V$.
\end{proof}
%Thus, Assumptions~\ref{assum:icat} and \ref{assum:involutive} hold.

\begin{propositionrep}\label{prop:theta-pos}
With the above definitions of $I$ and $\Phi$, a poset arrow $f\colon X \to \down Y$ is in correspondence with the relation $R \subseteq X \times Y$ satisfying:
$
\forall_{x,x'\in X,y,y'\in Y}\ (x\mathrel R y \land x \leq x' \land y' \leq y) \implies x' \mathrel R y' .
$
Moreover, Assumption~\ref{assum:Texistence} is valid.
\end{propositionrep}
\begin{proof}
Let $f\colon X\to \down Y$ and consider the relation $\theta_{X,Y}(f) = \{(x,y) \mid y \in fx\}$. Then it is straightforward to verify that it is upward closed in $X$ and downward closed in $Y$. Now consider a relation $R\subseteq X \times IY$ with this property and define the function $f\colon X \to \down Y$ as $x \mapsto \{y \mid x \mathrel R y\}$. Clearly, this mapping is well defined since $R$ is downward closed in $Y$. Moreover, if $x \leq x'$, then $fx \subseteq fx'$ because
\[x \mathrel R y \land x\leq x' \implies x' \mathrel R y.\]
Let $f\colon X \to X'$ and $h\colon X' \to TY$.  Then we find
\begin{align*}
(f\times IY)^* \theta_{X',Y}h =&\ (f\times IY)^* \{(x',y) \mid y\in hx'\}\\
 =&\ \{(x,\bar y) \mid (f\times IY)(x,\bar y) \in \{(x',y) \mid y\in hx'\}\}\\
 =&\ \{(x,y) \mid y \in h(fx)\} = \theta_{X,Y} f^*h.
\end{align*}
Let $g\colon Y \to Y'$ and $k\colon X\to TY$. Then, we find
\begin{align*}
\exists_{X\times Ig} \theta_{X,Y} (k) =&\ \exists_{X\times Ig} \{(x,y) \mid y\in k x\}\\
=&\ \mathord\uparrow \{(x,gy) \mid y\in kx \}\\
=&\ \{(\bar x,y') \in X \times IY' \mid x\leq \bar x \land y' \leq gy \land y\in kx\}\\
=&\ \{(\bar x,y') \in X \times IY' \mid x\leq \bar x \land y' \in \history g(kx)\}\\
=&\ \{(x,y') \mid y'\in \history g(kx)\} = \theta_{X,Y'} (Tg \circ k) .
\end{align*}
Note that the penultimate equality holds since. $\theta_{X,Y}(\_)$ produces a relation that is upward closed in the first argument. Thus, Assumption~\ref{assum:Texistence} holds.
\end{proof}
We take the same predicate lifting $\sigma$ as in the previous case since $\sigma_X(U) = \mathbb{A}\times U$ is upward closed whenever $U$ is upward closed. So, Assumption~\ref{assum:plift} is also valid. Now we again compute a relation lifting $\tilde\sigma_{X,Y} \colon \Phi(X \times IY) \to \Phi (AX \times IAY)$ to be (below $\uparrow$ denotes the upward closure of a subset):
{\allowdisplaybreaks
\begin{align*}
\tilde\sigma_{X,Y} R =&\ \exists_{\lambda_{X,IY}} \sigma_{X\times IY} R\\
=&\ \exists_{\lambda_{X,IY}}  \{(a,x,y) \mid x \mathrel R y \land a\in \mathbb{A}\}\\
=&\ \mathord{\uparrow} \{\lambda_{X,IY}(a,x,y) \mid x\mathrel R y \land a\in \mathbb{A}\}\\
=&\ \mathord{\uparrow} \big\{ \big( (a,x),(a,y)\big) \in AX \times IAY \mid x\mathrel R y \land a\in \mathbb{A} \big\}\\
=&\ \big\{ \big( (a,x),(a,y)\big) \in AX \times IAY \mid x\mathrel R y \land a\in \mathbb{A} \big\}.
\end{align*}
}
\begin{propositionrep}
In this setting, the `identity' relation $\Delta_X\subseteq X \times IX$ (for each $X\in\pos$) is given by $\{(x,x') \mid {x' \leq x} \land {x \in X}\}$. The `relational' composition $\odot$, in this setting, coincides with the usual relational composition of binary relations. Moreover, the relation lifting $\tilde\sigma_{X,Y}\colon \Phi (X\times IY) \to \Phi (AX \times IAY)$ preserves identity relations and relational compositions.
\end{propositionrep}
\begin{proof}
Recall that $\Delta_X = \theta_{X,X} (\eta_X)$, which results in the desired set $\{(x,x') \mid x' \leq x \land x \in X\}$. Moreover,
\[
\tilde \sigma_{X}(\Delta_X) =\Big\{ \big( (a,x),(a,x')\big) \in AX \times IAX \mid x\mathrel{\Delta_X} x' \land a\in \mathbb{A} \Big\},
\]
which is equivalent to the relation $\Delta_{AX}$ since the set $\mathbb{A}$ is discretely ordered.

Furthermore, the relation lifting preserves relational composition $\odot$ if we can show that $\odot$ coincides with the relational composition. To this end, consider two Kleisli arrows $f\colon X \to Y,g\colon Y \to Z$ in $\klcat(\down)$ and their composition $g\bullet f$:
\[
g\bullet f x = \mu_Z \circ \history\{gy \mid y\in fx\} = \{z\mid \exists_{W,y}\ z\in W \land W \subseteq gy \land y\in fx \} = \{z\mid \exists_y\ z\in gy \land y\in fx\}.
\]
Thus, $\odot$ coincide with the relational composition (just like in the case of $\set$).
\end{proof}
Moreover, the relation lifting $\tilde\sigma$ on a component evaluates like in the case of $\set$, we obtain a \kllaw{} $\vartheta \colon A \down \Rightarrow \down A$ given by
$
\vartheta_X(a,U) = \{(a,x) \mid x\in U\}
$,
for each $X\in \pos$. So we get a Kleisli lifting $\overline A \colon \klcat (\down) \to \klcat (\down)$ which is defined exactly like in the previous case. Nevertheless, the Kleisli lifting $\widehat B\colon \klcat(\gmonad\down) \to \klcat(\gmonad\down)$ (due to \cref{thm:KlExtMachineEndo}) on an arrow $f\colon \klcat (\gmonad \down)(X,Y)$ is a bit different and it evaluates as follows: $\widehat B(f)(k,o) = \history \{(k,o)\} $ and using \cref{eq:MachineEndo_def} we get
\begin{align*}
\widehat B(f)(k,a,x) = &\ TG\iota_{AY} \circ \widetilde Af (k,a,x) \\
=&\ TG\iota_{AY} \{(k',(a,y)) \mid (k',y)\in f(k,x)\}\\
=&\ \history { \big\{ \big( k',(a,y) \big) \mid (k',y) \in f(k,x) \big\} }\\
=&\ \big\{ \big( k',(a,y) \big) \mid (k',y) \in f(k,x) \big\}.
\end{align*}
Note that $\klcat(\down)$ is $\cppo$-enriched by taking the pointwise order, i.e.\ $f\leq g$ iff $\forall_{x\in X}\ fx \subseteq gx$ for any $f,g\colon X\to Y \in \pos$, and the join is given by the union (since it preserves downward closed subsets).
\begin{propositionrep}
The Kleisli lifting $\overline A$ is locally continuous and the operation $\_ + f$ (for every $f\in\klcat (\down)$) on Kleisli homsets commutes with $\omega$-directed joins. Moreover, the initial algebra of $B$ is $\fseq{\mathbb{A}}\times O$. As a result, the initial algebra $L(\mu_B) = \fseq{\mathbb{A}} \times O$ coincides with the final coalgebra of $\widehat B$ in $\klcat (\gmonad\down)$.
\end{propositionrep}
\begin{proof}
The proof that $\overline A$ is locally continuous and the operation $\_ + f$ commutes with $\omega$-directed joins is similar to \cref{prop:Set-CTS-coincide}. Below we argue why $\fseq{\mathbb{A}}\times O$ is the initial algebra for $B$ in $\pos$. The coincidence of initial algebra and final coalgebra follows from \ref{thm:finalcoal-machine}.

For the carrier set $\fseq{\mathbb{A}}\times O$ of the initial algebra we use the mapping $$\beta\colon \mathbb{A}\times \fseq{\mathbb{A}}\times O+O\rightarrow \fseq{\mathbb{A}}\times O\qquad (a,w,o)\mapsto(aw,o)\quad
o\mapsto(\epsilon, o)$$

If we work in \pos, we also have to argue this is order preserving. If $o'\leq o$, then $(\epsilon, o')\leq(\epsilon, o)$ and, since we trivially order the alphabet $\mathbb{A}$ and $\fseq{\mathbb{A}}$, $(a,w,o')\leq(a,w,o) \Rightarrow (aw,o')\leq(a,w,o')$.

Then for set $X$ and algebra $\alpha\colon \mathbb{A}\times X+O\rightarrow X$ we define the algebra homomorphism $f\colon \fseq{\mathbb{A}}\times O\rightarrow X$ as:
$$f(aw,o)=\alpha(a, f(w,o))\qquad f(\epsilon, o)=\alpha(o).$$
Note that $f$ is order preserving by induction on $w$.

We now have to prove that $\alpha\circ Bf = f\circ \beta$; for that purpose let $a\in \mathbb{A}$, $w\in \fseq{\mathbb{A}}$, $o\in O$, then:
$$f\circ\beta(a,w,o)=f(aw,o)=\alpha(a,f(w,o))$$
and
$$\alpha\circ Bf(a,w,o)=\alpha\circ(\mathbb{A}\times f+O)(a,w,o)=\alpha(a,f(w,o)).$$
For the case of just an element $o\in O$ we compute:
$$f\circ\beta(o)=f(\epsilon, o)=\alpha(o) \qquad \alpha\circ Bf(o)=\alpha\circ(\mathbb{A}\times f+O)(o)=\alpha(o).$$
\end{proof}
\newpage
\hbnote{Remove the failure equivalence from this theorem and add a remark why failure pairs can't be modelled this way.}
\begin{theoremrep}
Let $\alpha\colon X \to AX + O \in\klcat(\gmonad \down)$ be a coalgebra.
\begin{enumerate}
\item If $O=1$ then the unique coalgebra homomorphism is given by  $(k,x)\mapsto \bigcup_{k'\leq k} \{k'\}\times L(x,k')$.
\item If $O=\power (\mathbb{A})$ and $\alpha$ models the ready set of a state, i.e.\ $(k',o)\in \alpha(k,x)$ iff $o$ is the set of actions enabled from the state $x$ at condition $k'\leq k$, then the unique coalgebra homomorphism is given by $(k,x) \mapsto \bigcup_{k'\leq k} \{k'\} \times R(x,k')$.
\end{enumerate}
\end{theoremrep}

\begin{proof}
We again take a look at language equivalence before turning towards failure equivalence. We first need to define the behaviour map $f$ according to
$$f(k,x)=\{(k', w)\mid k'\leq k \land w\in L(x,k')\}$$

If we work in \pos\ we need to argue that the map is order preserving: Since $X$ is trivially ordered, we only need to consider the order on $\condset$; if $k'\leq k$, then $f(k',x)\subseteq f(k,x)$ by definition of $f(k,x)$.

Then we can compute:
\begin{equation*}\begin{aligned}
&\mu\circ\power \widehat Bf\circ\alpha(k,x) = \mu\power \widehat Bf(\{(k',\bullet),(k',a,x')\mid k\leq k \land x\xrightarrow{a,k}x'\})\\=&\{(k',\bullet),(k',a,w)\mid \exists x'\in X: w\in
L(x',k')\wedge
x\xrightarrow{a,k'}x'\wedge k'\leq k\}
\end{aligned}\end{equation*}
and:
\begin{equation*}\begin{aligned}
&\mu\circ\power \beta \circ f(k,x) =  \mu\power \beta (\{(k',w)\mid w\in L(x,k'), k'\leq k\})\\
=&\{(k',\bullet)\mid \epsilon\in L(x,k'), k'\leq k\}\cup\{(k',a,w)\mid aw\in L(x,k', k'\leq k)\}\\
=&\{(k',\bullet),(k',a,w)\mid \exists x'\in X: w\in L(x',k')\wedge x\xrightarrow{a,k}x'\wedge k'\leq k\}
\end{aligned}\end{equation*}

So the coalgebra homomorphism diagram commutes, concluding our proof for language equivalence.

For $O=\power (\mathbb{A})$, we consider the following map ($\condset \times X \to \power (\condset \times A^* \times \power (\mathbb A))$): $$f\colon X\rightarrow \widehat B(\fseq{\mathbb{A}}\times\power (\mathbb{A}))\qquad f(k,x)=\bigcup\{\{k'\}\times R(x,k')\mid k'\leq k\}$$where $R(x,k)$
for any state $x\in X$ and any condition $k\in K$ is the set of ready pairs starting in $x$ under the condition $k$.

To prove that the image of $f$ is downwards closed, let $(k',w,U) \in f(k,x)$ and $k'' \leq k'$ then $(k'',w,U) \in f(k,x)$ because upgrades preserve all active actions (but may add additional ones).

We also have to show that $f$ is order preserving, for that purpose, let $k' \leq k$, then $f(k',x) \subseteq f(k,x)$ trivially by definition, because $f$ filters by the condition downwards.

If we work in \pos\ we need to argue that the map is order preserving: Since $X$ is trivially ordered, we only need to consider the order on $K$; if $k'\leq k$, then $f(k',x)\subseteq f(k,x)$ by definition of $f(k,x)$.

We now have to show that $\alpha$ maps to a downwards closed set, so let $(k',U) \in \alpha(k,x)$ and $k''\leq k'$ then $(k'',U) \in \alpha(k,x)$, because upgrades preserve all transitions and thus all ready actions.

Clearly, two states map to the same point in the final coalgebra iff they
are ready equivalent, so we we only need to show that the arrow $f$ is actually a coalgebra homomorphism (therefore necessarily unique):
\begin{equation*}\begin{aligned}
&\mu\circ\power \widehat Bf\circ\alpha(k,x) =  \mu\power \widehat Bf(\{(k',a,x')\mid x\xrightarrow{a,k}x', k'\leq k\}\cup\{(k',U)\mid\forall a\in U\exists x'\in X: x\xrightarrow{a,k'}x'\wedge k'\leq
k\})\\=&\bigcup\{\{k''\}\times\{a\}\times F(x',k'')\mid x\xrightarrow{a,k}x'\wedge k''\leq k'\leq k\}\cup\{(k',U)\mid \forall a\in U\exists x'\in X: x\xrightarrow{a,k'}x'\wedge k'\leq k\}
\end{aligned}\end{equation*}
and:
\begin{equation*}\begin{aligned}
&\mu\circ\power \beta \circ f(k,x) = \mu\power \beta (\bigcup\{\{k'\}\times R(x,k')\mid k'\leq k\})\\
=&\begin{multlined}[t]
    \{(k',U)\mid (\epsilon,U)\in R(x,k')\wedge k'\leq k\} \\ \cup\{(k'',a,(w,U))\mid\exists x'\in X:x\xrightarrow{a,k'}x'\wedge (w,U)\in F(x',k'')\wedge k''\leq k'\leq k\}
\end{multlined}\\
=&\{(k',U)\mid \forall a\in U\exists x'\in X: x\xrightarrow{a,k'}x'\wedge k'\leq k\}\cup\bigcup\{\{k''\}\times\{a\}\times F(x',k'')\mid x\xrightarrow{a,k'}x'\wedge k''\leq k'\leq k\}
\end{aligned}\end{equation*}

Note that we can characterise failure equivalence in the presence of upgrades completely analogously in \set: We only have to change the map $f$ to:
$$f\colon X\rightarrow \widehat B(\fseq{\mathbb{A}}\times\power (\mathbb{A}))\qquad f(k,x)=\bigcup\{\{k'\}\times F(x,k')\mid k'\leq k\}$$where $F(x,k)$
for any state $x\in X$ and any condition $k\in K$ is the set of failure pairs starting in $x$ under the condition $k$. The computation is completely analogous. However, note that this is not a suitable choice in \pos, because $\alpha$ is not downwards closed and neither is the final coalgebra map.
\end{proof}

\section{Conclusion and future work}

In this paper, we developed a coalgebraic framework to describe CTSs with and without upgrades that allowed us to synthesise language, failure and ready equivalences through our main theorem, \cref{thm:finalcoal-machine}.
The crucial assumption for this main theorem turned out to be the lifting property to the Kleisli category, which is equivalent to defining a \kllaw{}.
Under certain assumptions, we characterised a \kllaw{} internally living in the fiber of an indexed category. We demonstrated how these assumptions can be easily checked in both cases---with and without upgrades.

Based on the development in \cref{sec:Kl-lift}, %one should conduct further case studies related to CTSs by considering relative monad from sub-distribution monad. In addition,
perhaps in the future, it would be worthwhile to investigate the sufficient conditions that guarantee the existence of initial algebra and final coalgebra for the Kleisli lifting $\overline F\colon \klcat (T) \to \klcat(T)$ constructed via the \kllaw{} $\vartheta\colon FT \to TF$. Moreover, if both exist, a question of particular interest will be whether they coincide. A major contribution would be, if one could attack this question by weakening the assumption of a celebrated result to this effect by Freyd~\cite{Freyd_1992} for $\cppo$-enriched categories, a set-up we alluded to.
To this end, it would be also fruitful to seek conditions when the Kleisli lifting constructed in \cref{eq:MachineEndo_def} is actually locally continuous.

Another fruitful direction would be to develop coinductive characterisation of failure equivalence in the presence of upgrades and consider other monads than the powerset monad in modelling quantitative extensions of CTSs.
Such an extension may enrich the transitions with real-time \cite{Cordy_2012_behavioural}, probability \cite{Rodrigues_2015_modeling} or even weights from a semiring modelling some resource usage \cite{olaechea_2016long,ter2015statistical}.
Probabilistic extensions are among the upmost exciting one---specially\ so-called ``parametric Markov models''---see \cite[Def.~3.6]{Arming2018_parameter}, and also \cite{Junges2024_parameter}---in which one considers the distribution monad and which originated from the formal verification of probabilistic systems.

\bibliographystyle{entics}
\bibliography{refer}
\end{document}